\documentclass[a4paper,12pt,twoside]{article}

\usepackage{psfig}
\usepackage{cite}
\usepackage[english]{babel}
\usepackage{graphics}
\usepackage{epsfig}
\usepackage{verbatim}
\usepackage{rotating}
\usepackage[centertags]{amsmath}
\usepackage{array}
\usepackage{multirow}
\usepackage{supertabular}
\usepackage{dcolumn}
\usepackage{mcite}
\usepackage{xspace}
\usepackage{graphicx}
\usepackage{amsfonts}
\usepackage{amssymb}

\input zeuspap_def.sty
\input zeuspap_fmt.sty

\def\xgo{x_{\gamma}^{\mbox{\rm\tiny OBS}}}
\def\xpo{x_p^{\mbox{\rm\tiny OBS}}} 
\def\xpio{x_{\pi}^{\mbox{\rm\tiny OBS}}} 
\def\xg{x_\gamma} 

\def\xl{x_{\mbox{\rm\tiny L}}}
\def\pt{p_{\mbox{\rm\tiny T}}}
\def\ETJ{E_{\mbox{\rm\tiny T}}^{\mbox{\rm\tiny jet}}}
\def\ETAJ{\eta^{\mbox{\rm\tiny jet}}}

\newcommand{\lap}
{\ensuremath{\stackrel{_{\scriptstyle <}}{_{\scriptstyle\sim}}}}

\newcommand{\lsim} {\mbox{\raisebox{-0.4ex}
{$\;\stackrel{<}{\scriptstyle \sim}\;$}}}

\newcommand{\Journal}[4]{{#1}{#2} (#3) #4}

\newcommand{\CPC}{Comp.\ Phys.\ Comm.\ }
\newcommand{\EPA}{Eur.\ Phys.\ J.\ A\ }
\newcommand{\EPC}{Eur.\ Phys.\ J.\ C\ }

\newcommand{\NIMA}{Nucl.\ Instr.\ and Meth.\ A\ }

\newcommand{\NPB}{Nucl.\ Phys.\ B\ }
\newcommand{\NPPS}{Nucl.\ Phys.\ Proc.\ Suppl.\ }

\newcommand{\PLB}{Phys.\ Lett.\ B\ }

\newcommand{\PRep}{Phys.\ Rep.\ }
\newcommand{\PRL}{Phys.\ Rev.\ Lett.\ }

\newcommand{\PRC}{Phys.\ Rev.\ C\ }
\newcommand{\PRD}{Phys.\ Rev.\ D\ }

\newcommand{\SNP}{Sov.\ J.\ Nucl.\ Phys.\ }
\newcommand{\ZPA}{Z.\ Phys.\ A\ }
\newcommand{\ZPC}{Z.\ Phys.\ C\ }

\begin{document}

\title{
\bf\LARGE  Measurement of dijet cross sections \\
           for events with a leading neutron\\
           in photoproduction at HERA \\ 
}

\author{ZEUS Collaboration}

\date{}

\maketitle

\begin{abstract} 
Differential cross sections for dijet photoproduction in
association with a leading neutron using the reaction
$e^+ + p \rightarrow e^+ + n + \mathrm{jet} + \mathrm{jet} + X_r$
have been measured with the ZEUS detector at HERA 
using an integrated luminosity of 6.4 pb$^{-1}$.
The fraction of dijet events with a leading neutron in the final state 
was studied as a function of the jet kinematic variables.
The cross sections were measured 
for jet transverse energies
\mbox{$\ETJ  > 6$ GeV},
neutron energy \mbox{$E_n > 400$ GeV},
and neutron production angle $\theta_n < 0.8$ mrad.
The data are broadly consistent with factorization of the
lepton and hadron vertices and with a simple
one-pion-exchange model.
\end{abstract} 

\vspace{-14.5cm}
\begin{flushleft}
\tt DESY 00-142 \\
October 2000 \\
\end{flushleft}

{\hspace*{-0mm} } \pagestyle{plain} \thispagestyle{empty} \newpage 

\topmargin-1.cm                                                                                    
\evensidemargin-0.3cm                                                                              
\oddsidemargin-0.3cm                                                                               
\textwidth 16.cm                                                                                   
\textheight 680pt                                                                                  
\parindent0.cm                                                                                     
\parskip0.3cm plus0.05cm minus0.05cm                                                               
\def\3{\ss}                                                                                        
\newcommand{\address}{ }                                                                           
\pagenumbering{Roman}                                                                              

                                                   %
\begin{center}                                                                                     
{                      \Large  The ZEUS Collaboration              }                               
\end{center}                                                                                       
  J.~Breitweg,                                                                                     
  S.~Chekanov,                                                                                     
  M.~Derrick,                                                                                      
  D.~Krakauer,                                                                                     
  S.~Magill,                                                                                       
  B.~Musgrave,                                                                                     
  A.~Pellegrino,                                                                                   
  J.~Repond,                                                                                       
  R.~Stanek,                                                                                       
  R.~Yoshida\\                                                                                     
 {\it Argonne National Laboratory, Argonne, IL, USA}~$^{p}$
% ...
\newpage                                                                                           
$^{\    1}$ now visiting scientist at DESY \\                                                      
% ...
$^{\    6}$ now at Siemens ICN, Berlin, Germanny \\                                                
% ...
                                                           %
                                                           %
\newpage   
                                                           %
                                                           %
\begin{tabular}[h]{rp{14cm}}                                                                       
$^{a}$ &  supported by the Natural Sciences and Engineering Research                               
          Council of Canada (NSERC)  \\                                                            
$^{b}$ &  supported by the FCAR of Qu\'ebec, Canada  \\                                            
$^{c}$ &  supported by the German Federal Ministry for Education and                               
          Science, Research and Technology (BMBF), under contract                                  
          numbers 057BN19P, 057FR19P, 057HH19P, 057HH29P, 057SI75I \\                              
$^{d}$ &  supported by the MINERVA Gesellschaft f\"ur Forschung GmbH, the                          
          Israel Science Foundation, the U.S.-Israel Binational Science                            
          Foundation, the Israel Ministry of Science and the Benozyio Center                       
          for High Energy Physics\\                                                                
$^{e}$ &  supported by the German-Israeli Foundation, the Israel Science                           
          Foundation, the U.S.-Israel Binational Science Foundation, and by                        
          the Israel Ministry of Science \\                                                        
$^{f}$ &  supported by the Italian National Institute for Nuclear Physics                          
          (INFN) \\                                                                                
$^{g}$ &  supported by the Japanese Ministry of Education, Science and                             
          Culture (the Monbusho) and its grants for Scientific Research \\                         
$^{h}$ &  supported by the Korean Ministry of Education and Korea Science                          
          and Engineering Foundation  \\                                                           
$^{i}$ &  supported by the Netherlands Foundation for Research on                                  
          Matter (FOM) \\                                                                          
$^{j}$ &  supported by the Polish State Committee for Scientific Research,                         
          grant No. 112/E-356/SPUB/DESY/P03/DZ 3/99, 620/E-77/SPUB/DESY/P-03/                      
          DZ 1/99, 2P03B03216, 2P03B04616, 2P03B03517, and by the German                           
          Federal Ministry of Education and Science, Research and Technology (BMBF)\\              
$^{k}$ &  supported by the Polish State Committee for Scientific                                   
          Research (grant No. 2P03B08614 and 2P03B06116) \\                                        
$^{l}$ &  partially supported by the German Federal Ministry for                                   
          Education and Science, Research and Technology (BMBF)  \\                                
$^{m}$ &  supported by the Fund for Fundamental Research of Russian Ministry                       
          for Science and Edu\-cation and by the German Federal Ministry for                       
          Education and Science, Research and Technology (BMBF) \\                                 
$^{n}$ &  supported by the Spanish Ministry of Education                                           
          and Science through funds provided by CICYT \\                                           
$^{o}$ &  supported by the Particle Physics and                                                    
          Astronomy Research Council \\                                                            
$^{p}$ &  supported by the US Department of Energy \\                                              
$^{q}$ &  supported by the US National Science Foundation                                          
\end{tabular}                                                                                      
                                                           %
                                                           %
\newpage

\newpage

\topmargin-1.5cm
\evensidemargin-0.3cm
\oddsidemargin-0.3cm
\textwidth 16.cm
\textheight 650pt
\parindent0.cm
\parskip0.3cm plus0.05cm minus0.05cm

\pagenumbering{arabic} 
\setcounter{page}{1}

\section{Introduction} 

A wealth of data\cite{erwin,pickup,engler,robinson,flauger,hanlon,hartner,eisenberg,blobel,abramowicz}
exists on charge-exchange processes in soft hadronic
reactions, where an initial-state proton is transformed into a 
final-state neutron, $p\rightarrow n$. A successful phenomenological
description of these results has been obtained with the concept of
the exchange of virtual isovector mesons, such as
$\pi$, $\rho$, and $a_2$, using
Regge theory\cite{yukawa,sullivan,bishari,field,ganguli,zakharov,zoller}. 
Since the pion is by far the lightest hadron, its
exchange dominates the $p\rightarrow n$ transition,
particularly at small values of the
squared momentum transfer, $t$, between the proton and the neutron.

The assumption of factorization, namely that the partonic nature of
a hadron is independent of the hard scattering process in which it
participates, has been shown to be valid in the
case of the nucleon, whose partonic structure has been probed
extensively in jet-production processes as well as
in deep inelastic scattering.
The idea of factorization may be extended to the exchanged
objects in charge-exchange reactions.
Under this assumption,
hard processes occurring in charge-exchange reactions, such as the
production of high-$E_T$ jets, 
provide a means of investigating
the partonic nature of the exchanged objects.

Charge-exchange processes have been studied in deep inelastic
scattering at HERA\cite{fnc2,h1f2lb}.
This paper reports the first observation of the photoproduction of
dijets in association
with an energetic forward neutron:
\begin{equation}
e^+ + p \rightarrow e^+ + n + \mathrm{jet}+ \mathrm{jet}+ X_r 
\label{eqn:taggedep}
\end{equation}
where $X_r$ denotes the remainder of the final state.
The virtuality of the exchanged photon, $Q^2$, was less 
than $\sim 4$~GeV$^2$, 
with a median value of about $10^{-3}$~GeV$^2$.
Neutrons with energy $E_n > 400$ GeV and
produced at an angle $\theta_n < 0.8$ mrad with respect to the 
direction\footnote{
The ZEUS coordinate system is defined as right-handed
with the $Z$ axis pointing in the proton 
beam direction, hereafter referred to as forward, 
and the $X$ axis horizontal, 
pointing towards the center of HERA.
Pseudorapidity is defined as 
$\eta=-\ln\left(\tan(\theta/2)\right)$, 
where the polar angle $\theta$ is taken 
with respect to the proton beam direction.
}
of the HERA proton beam ($E_p=820$ GeV)
were detected in a forward neutron calorimeter.
These results extend previous ZEUS photoproduction 
dijet studies\cite{zeusdij95,dijet_98,zeus_dijet_diff}.

The present data are compared to an inclusive sample of
dijet events selected without the requirement of a forward neutron.
Cross sections are presented both as a function of
the kinematic variables of the jet and of the
$p\rightarrow n$ transition.
The contributions to the cross section of direct processes, 
where the photon acts
as a point particle, and resolved, where the photon
acts as a source of partons, are compared in 
both the neutron-tagged and the inclusive samples.
In addition, the fraction
of the inclusive 
dijet sample with a leading neutron 
is given as a function of the
jet transverse energy ($\ETJ$) and pseudorapidity ($\ETAJ$). 
These fractions are used to study 
the factorization properties of the processes. 
The results are compared to predictions of one-pion exchange. 



\section{Event kinematics}
\label{sec:kine}

The dijet processes under consideration here are characterized by
an initial state consisting of a positron $e^+$, and a proton $p$, and
a final state consisting of the scattered
positron, the scattered neutron $n$, and a hadronic system $H$:
\begin{equation}
  e^+(k) + p(P) \rightarrow e^+(k') + n(P') + H
\end{equation}
where $k$, $k'$, $P$ and $P'$ are the four-momenta
of the initial and scattered positron, and the proton and
neutron, respectively. 
The process is described by four Lorentz invariants.
Two describe the positron-photon vertex and can
be taken to be the virtuality of the exchanged photon ($Q^2$) and
the electron's inelasticity ($y$), defined by:
\begin{eqnarray}
   Q^2&=& -q^2 = -(k-k')^2\\
     y&=&\frac{P\cdot q}{P\cdot k}
\end{eqnarray}
In photoproduction, where $Q^2$ is small,  
$y=(E_e-E')/E_e=E_\gamma/E_e$, where $E_e(E')$ is the energy
of the initial (scattered) positron, and $E_\gamma$ is the
incident photon energy.
The other two variables, which describe the proton-neutron vertex, 
are the fraction of the energy of the initial-state proton
carried by the
neutron ($\xl$), and the square of the momentum transfer ($t$)
between the initial proton and the produced neutron,
defined by:
\begin{eqnarray}
   \xl&=&\frac{P'\cdot k}{P\cdot k}\simeq \frac{E_n}{E_p}\\
     t&=&(q')^{\;2}=(P-P')^2
\end{eqnarray}
where $E_p$ is the energy of the incident proton.
The transverse momentum, $\pt$, of the neutron is related to 
$t$ and $\xl$ by:
\begin{equation}
   t=-\frac{\pt^2}{\xl}-\frac{(1-\xl)(m_n^2-\xl m_p^2)}{\xl}
\label{eqn:t}
\end{equation}
where $m_p (m_n)$ is the mass of the proton (neutron).


In the photoproduction of dijets tagged with a leading neutron,
the hadronic system H contains
at least two jets:
\begin{equation}
  e^+(k) + p(P)\ \rightarrow\  e^+(k') + n(P') + H
               \ \rightarrow\  e^+(k') + n(P') + {\rm jet+jet}+X_r
\label{eqn:gpn}
\end{equation}
In $2\rightarrow 2$ scattering of
massless partons, the fractions
of the four-momenta $q=(k-k')$ and $q'=(P-P')$
carried into the hard scattering
by the initial-state partons
are given by:
\begin{eqnarray}
x_{\gamma}&=&\frac{(p_{J1}+p_{J2})\cdot q'}{q\cdot q'}\\ 
x_{\pi}   &=&\frac{(p_{J1}+p_{J2})\cdot q }{q\cdot q'}
\end{eqnarray}
respectively,
where $p_{Ji}$ is the four-momentum of the $i$th final-state parton,
and the approximation
$q^2\approx (q')^2\approx 0$ has been used.
The energy fraction contributed by the exchanged photon to the production of
the dijets is $x_{\gamma}$;
in Regge models, where $p\rightarrow n$ is
the result of the $\pi$, $\rho$ or $a_2$ trajectory 
coupling to the $pn$ vertex,
the corresponding contribution of the exchanged meson is $x_{\pi}$.
A further relationship is:
\begin{equation}
      x_{\pi}=x_p/(1-\xl )
\end{equation} 
where $x_p$ is the fraction of the proton energy participating
in the production of the dijets.

The observables
$\xgo$,  $\xpo$ and $\xpio$, 
defined in terms of jets,
are introduced\cite{zeusdij95}:
\begin{eqnarray}
\xgo &=& \frac {\sum_\mathrm{jets} \ETJ  e^{-\ETAJ}}{2 E_\gamma}\\
\xpo &=& \frac {\sum_\mathrm{jets} \ETJ  e^{\ETAJ}}{2 E_p}\\ 
\xpio &=& \frac {\xpo}{(1-\xl )}\label{eqn:xpi}
\end{eqnarray}
where the sum runs over the two jets of highest $\ETJ$ in an event.
The variables $\xgo$ and $\xpio$ are
estimators of $x_{\gamma}$ and $x_{\pi}$, respectively.

\begin{figure}[htb]
\includegraphics[width=\linewidth]{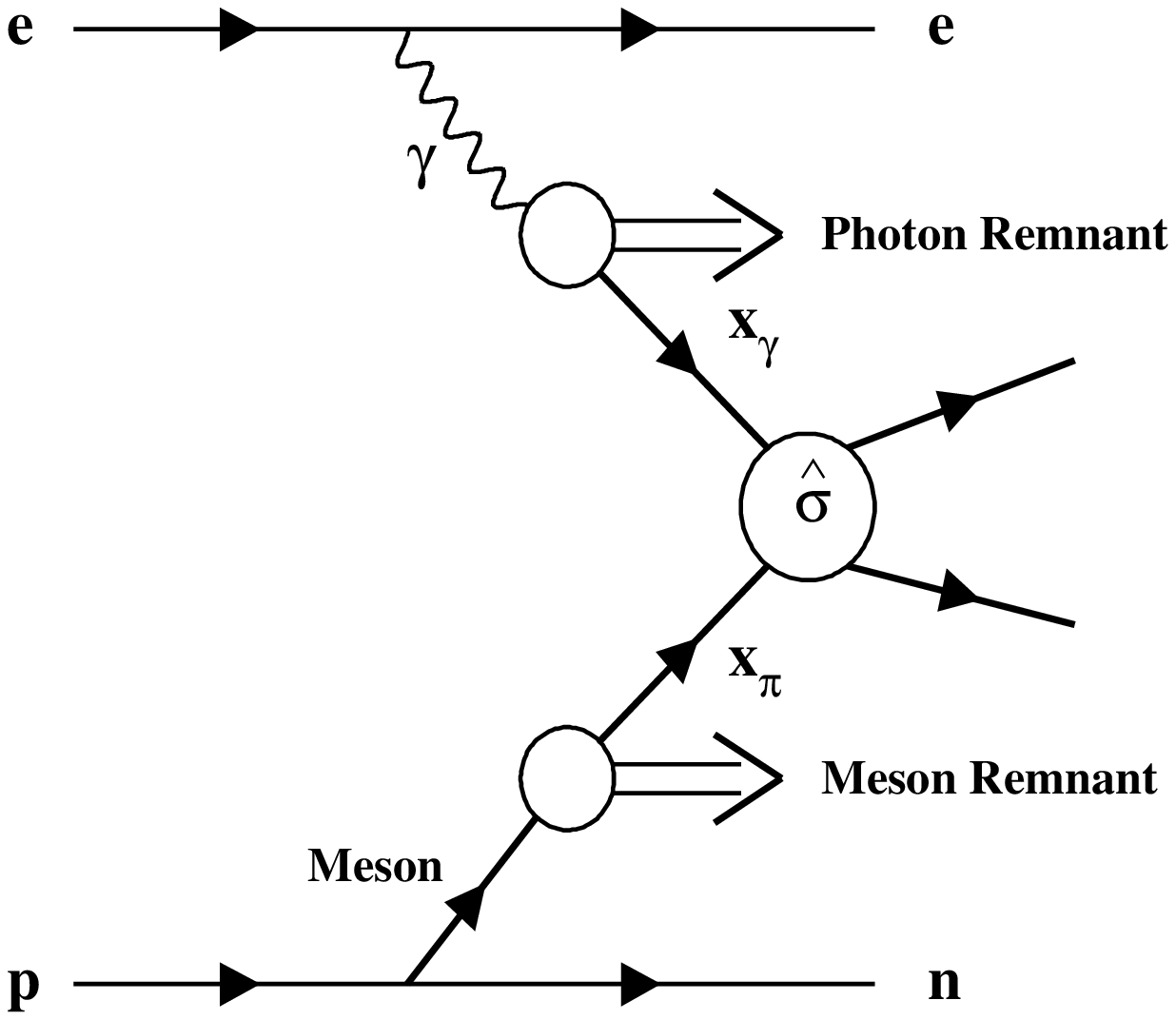}
\caption{ 
Resolved photoproduction of dijet events with a leading
neutron through meson exchange. The fraction of the 
energy of the exchanged meson (photon) 
participating in the partonic hard scattering
that produces the dijet system
is denoted by $x_{\pi}$ ($x_{\gamma}$);
the corresponding hard cross section is 
$\hat{\sigma}$. In direct photoproduction, the entire exchanged photon
participates in the hard scattering, there is no
photon remnant, and $x_{\gamma}=1$.}
\label{fig:feynman}
\end{figure}

In leading-order (LO) QCD, two types of
processes contribute to jet photoproduction \cite{owens,drees}: 
either the entire photon interacts with a parton in the target (the 
{\it direct} process), or the photon acts as a source of partons which scatter 
off those in the target (the {\it resolved} process). 
Figure~\ref{fig:feynman} illustrates these processes for
reaction~(\ref{eqn:gpn}) with an assumed meson-exchange
contribution.
The observable $\xgo$ is sensitive to 
which type of process occurs\cite{zeusnov93}. 
Direct processes are 
concentrated at high values of $\xgo$,
resolved processes at low values.



\section{Experimental conditions}

The data sample used in this analysis was collected in 1995
with the ZEUS detector using $e^+p$ interactions.
In this period HERA operated with 174 colliding bunches of $E_p=$~820~GeV 
protons and $E_e=$~27.5~GeV positrons,
corresponding to a center-of-momentum-frame energy
$\sqrt s = 300$~GeV.
Additionally, 21 unpaired 
bunches of protons or positrons allowed the 
beam--related backgrounds to be determined. 
The integrated luminosity used in this analysis is
\mbox{6.4 pb$^{-1}$}.

The ZEUS detector is
described in detail elsewhere\cite{zeusdet}. 
The principal components used in the present analysis were the central
tracking detector (CTD)\cite{ctd} 
positioned in a 1.43~T solenoidal magnetic field,
the uranium-scintillator sampling calorimeter (CAL)\cite{cal}, and the
forward neutron calorimeter (FNC)\cite{fnc3}.
The tracking system was used to establish 
an interaction vertex with a typical resolution 
along (transverse to) the beam direction of
0.4(0.1) cm.
Energy deposits in the CAL were used to find jets and to measure
their energies and angles.
The CAL is hermetic and consists of 5918 cells, each 
read out by two photomultiplier tubes. Under test beam conditions,
the CAL has energy resolutions of $\sigma(E)=18\%\sqrt E$
for electrons and $35\%\sqrt E$ for hadrons ($E$ in GeV)\cite{cal}.
Jet energies were corrected for the
energy lost in inactive material 
(typically one radiation length) 
in front of the CAL. 

\subsection{Forward neutron calorimeter}

\begin{figure}[htb]
\begin{minipage}[t]{5cm}
\centerline{\hspace{1cm}\psfig{file=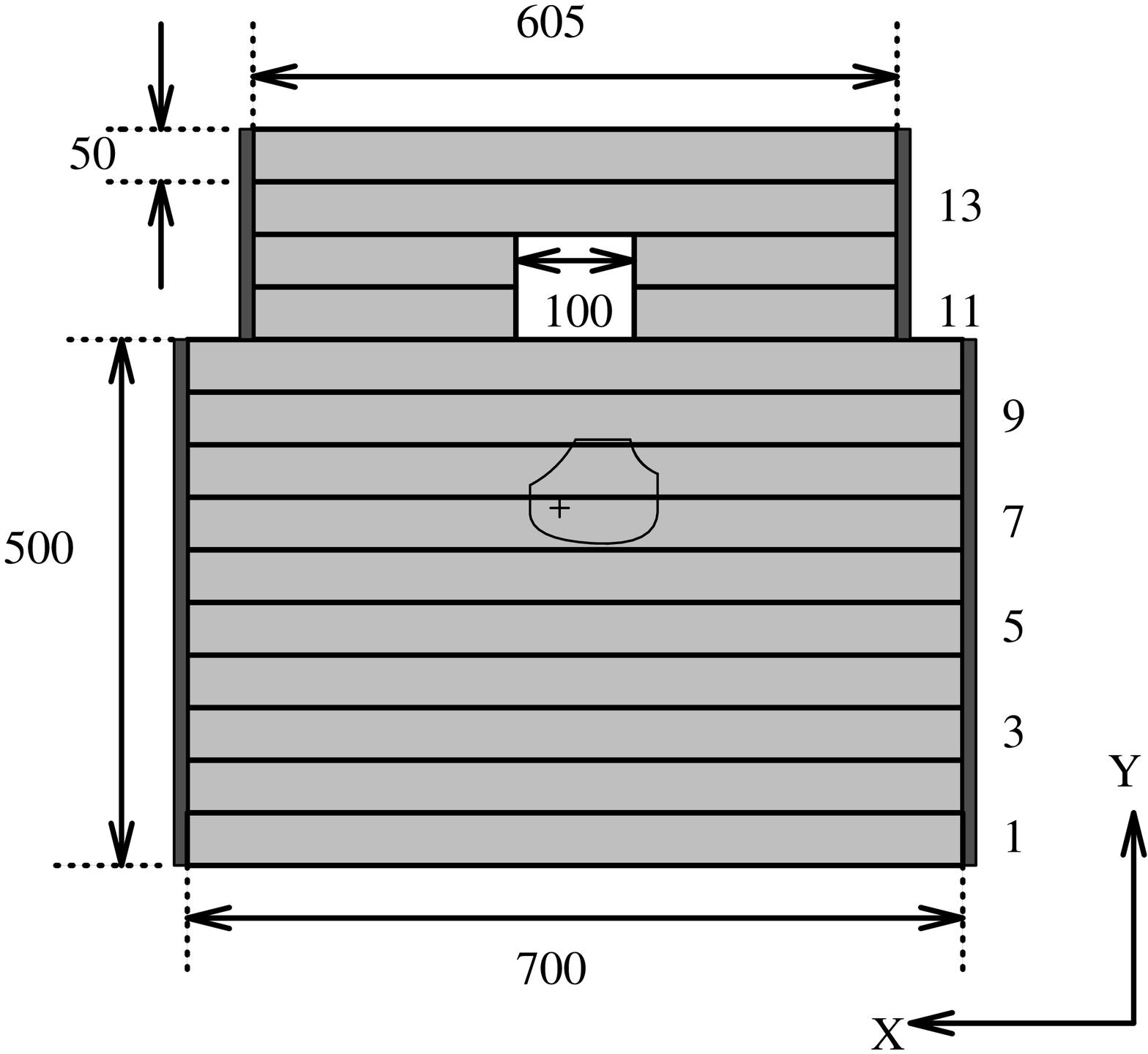,height=7.cm,width=7.cm}}
\centerline{(a)}
\end{minipage}\hspace{3cm}
\begin{minipage}[t]{7cm}
\centerline{\hspace{1cm}\psfig{file=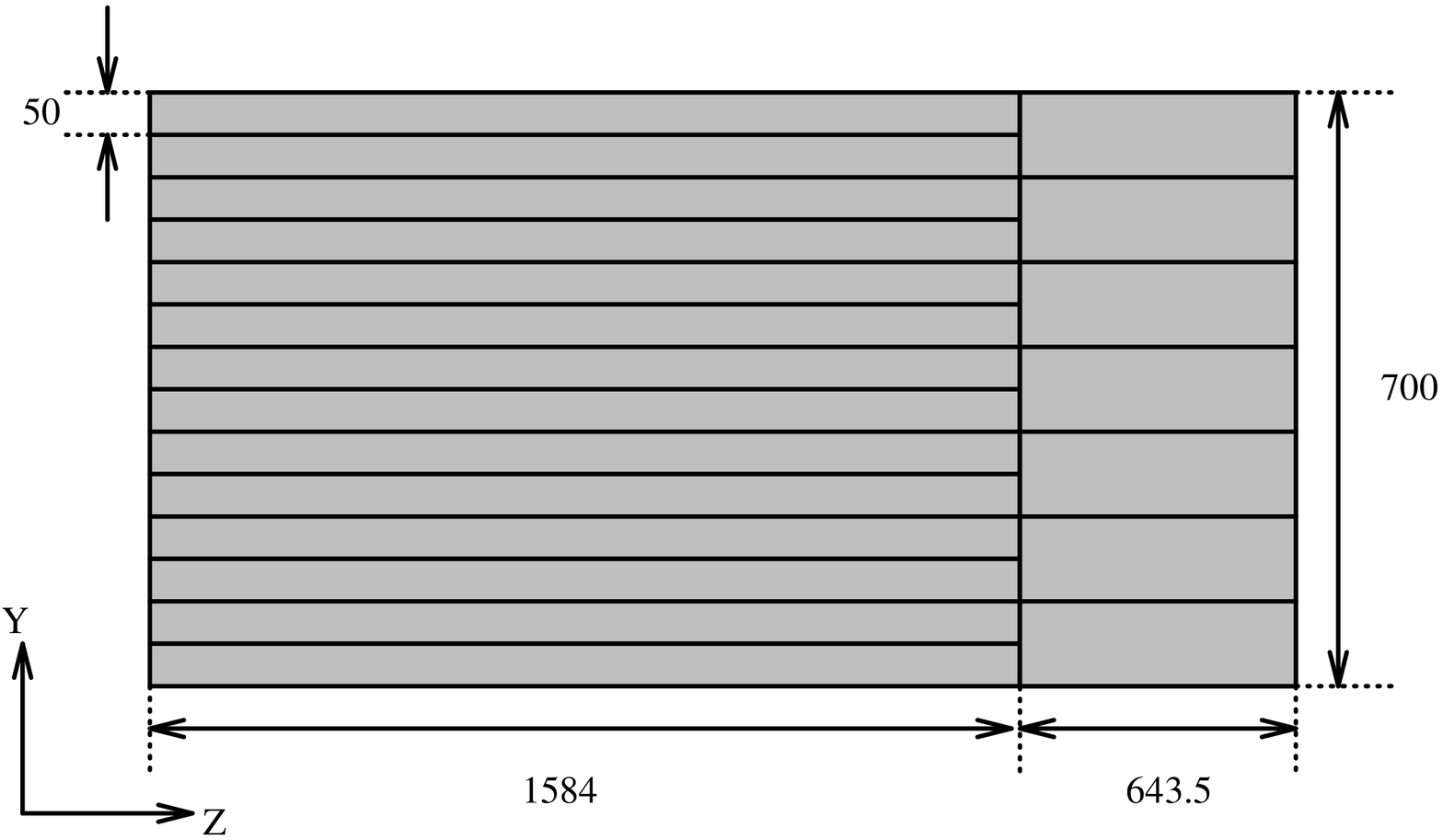,height=7.cm,width=10.cm}}
\centerline{\hspace{1.5cm}(b)}
\end{minipage}
\caption{
(a) Front view of the FNC showing the top
and bottom sections. The outline in towers 7 and 8 shows
the open geometric aperture for the incident neutrons. 
The point (+) marks zero degrees.
The hole in towers 11 and 12 accommodates
the proton beam pipe of HERA. 
The darker shading indicates
the location of the wavelength-shifting 
light-guides.
(b) A side view showing the front and rear sections. 
The vertical segmentation in the front is 50 mm; the
segmentation in the rear is 100 mm. 
All dimensions are in mm.}
{\label{fig:fnc}}
\end{figure}

The forward neutron calorimeter\cite{fnc3} 
was installed in the HERA tunnel at 
\mbox{$\theta = 0$} degrees 
and at $Z = 106$ m from the interaction point in the 
proton-beam direction, and used for the 1995 data taking.
The layout of the calorimeter is shown in 
Fig.~\ref{fig:fnc}. 
The FNC is a 
sampling calorimeter 
with 134 layers of 1.25~cm thick 
lead as the absorber and 0.26~cm thick scintillator as the active material.
The scintillator is read out on each side with
wavelength-shifting light-guides coupled 
to photomultiplier tubes. 
It is segmented longitudinally into a front section, seven 
interaction-lengths deep, and a rear section, three interaction-lengths deep.
The front section is divided vertically into 14 towers,
allowing the separation of electromagnetic and hadronic showers
using the energy-weighted vertical width of the showers.  
The energy resolution for hadrons,
as measured in a beam test, 
is
$65\%\sqrt{E}$ ($E$ in GeV)\cite{fnc3}. 
Neutrons are easily distinguished from protons,
which are deflected upwards by the beam magnets
and deposit most of their energy in
the top four towers of the FNC.

Three planes of veto counters, 
each $70\times 50\times\,2$ cm$^3$, 
are located 70, 78, and 199 cm in front
of the calorimeter.  
These counters, which completely cover the bottom front
face of the calorimeter, were used offline to identify
charged particles and so reject particles
which interacted in the inactive material
in front of the FNC. 

\begin{figure}[ht]
\includegraphics[width=\linewidth]{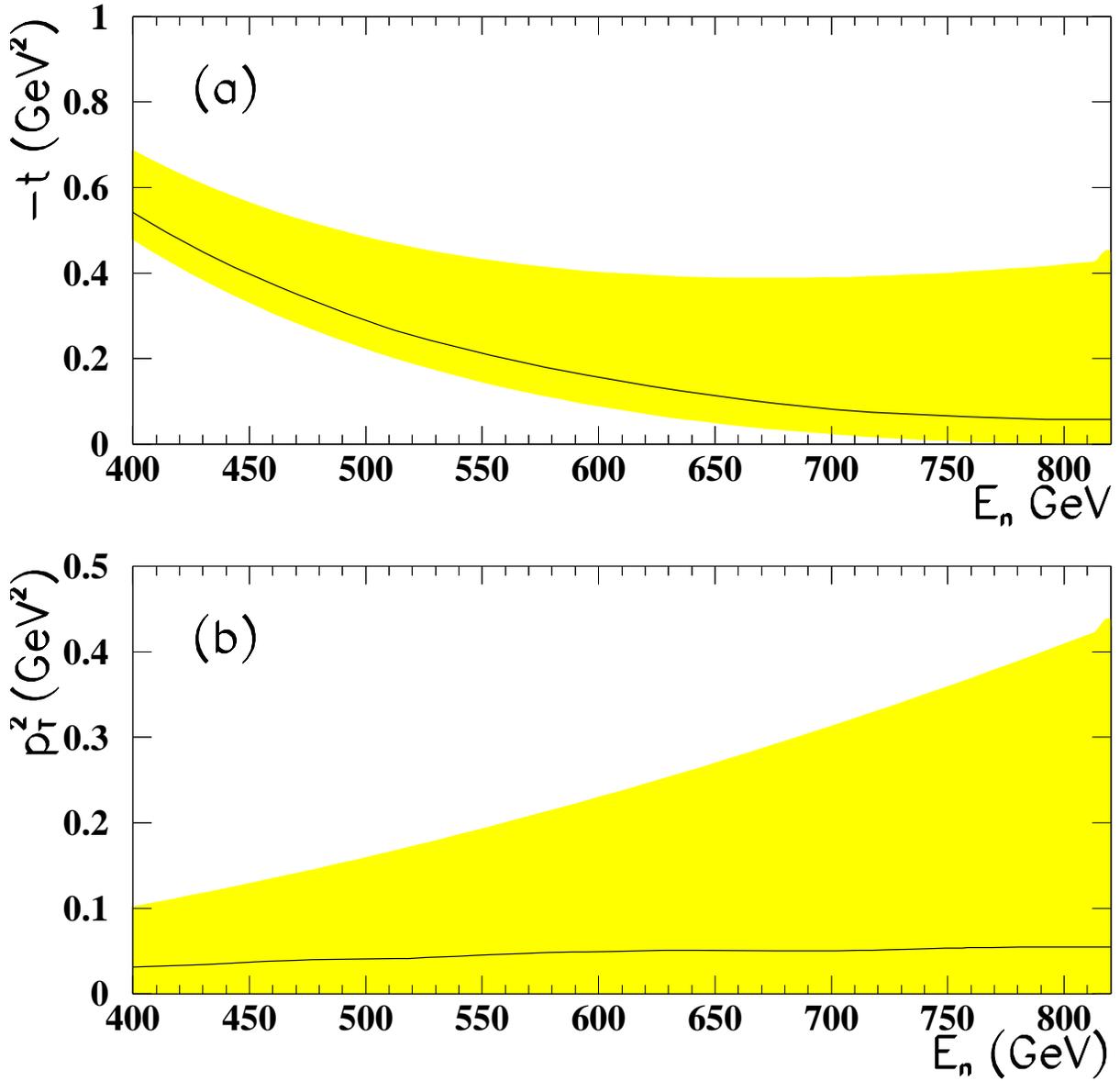}
\caption{ The kinematic regions in (a) $t$, and (b)
$\pt^2$ covered by the angular acceptance of the FNC
($\theta < 0.8$ mrad) are shown as shaded bands.
The solid lines show the average $t$
and $\pt^2$, respectively, as function of the neutron energy $E_n$.}
\label{fig:kine}
\end{figure}

Magnet apertures limit the FNC acceptance to neutrons with 
production angles less than \mbox{$0.8$ mrad},
that is to transverse
momenta $\pt\le E_n\theta_{\mbox{\rm\tiny max}}=0.66\xl$~GeV. 
Only about one quarter of the
corresponding azimuth is free of obstruction, as can be seen
from the outline of the aperture in Fig.~\ref{fig:fnc}(a).
The $Z$-axis intersection with the FNC is also indicated.
The overall acceptance of the FNC, which includes beam-line geometry, 
absorbing material, and the angular distribution of the neutrons,
is about 30\% 
for neutrons with energy \mbox{$E_n > 400$ GeV}
and \mbox{$\theta_n<0.8$ mrad}.
The kinematic region covered by the FNC in longitudinal and transverse
variables is shown in Fig.~\ref{fig:kine}. Although the
acceptance extends to $\pt^2\simeq 0.4$ GeV$^2$,
the mean value of $\pt^2$ for the accepted data
is less than 0.05 GeV$^2$\cite{fnt_osaka}. The $t$ acceptance 
is strongly affected by the 
minimum value of $|t|$,
$t_{\mbox{\rm\tiny min}}=(1-\xl)(m_n^2-\xl m_p^2)/\xl$.

The calibration and monitoring\cite{calor97} of the FNC
follow the methods developed for the FNC test
calorimeters\cite{fnc2,fnc1}. The gain of the 
photomultiplier tubes is
obtained by scanning the FNC with a $^{60}$Co 
radioactive source. Changes in gain during data taking are monitored 
using energy deposits from interactions of the HERA proton beam with
residual gas in the beam pipe.
The overall energy scale is set from the kinematic end point
of 820 GeV by fitting proton beam-gas interaction data with
energy greater than 600 GeV
to that expected from one-pion exchange\cite{bishari,holtmann}. 


\section{Data selection}
\label{sec:ev-sel}

The ZEUS detector uses 
a three-level trigger system.  At the first level,
events were selected by a coincidence of a regional
or transverse energy sum in the calorimeter, and at least one track
from the interaction point measured in the CTD.
At the second level, at least 8~GeV total transverse energy, excluding
the eight calorimeter towers immediately surrounding the forward
beam pipe, was required and cuts on calorimeter energies and timing were
used to suppress events caused by beam-gas interactions~\cite{F2}.
At the third level, a
cone algorithm used the calorimeter cell energies
and positions to identify jets. Events were required to have at least
two jets, each of which satisfied the requirement
$\ETJ > 3.5$~GeV and $\ETAJ < 2.0$, or
$\ETJ > 4.0$~GeV and $2.0 < \ETAJ < 2.5$.  Additional tracking cuts
were made to reject proton beam-gas interactions and cosmic-ray
events.
No requirement on the FNC was made at any trigger level. 

Further selection criteria were applied offline. 
Charged current
scattering events were rejected by a cut on the missing transverse
momentum measured in the calorimeter. To reject remaining beam-gas and
cosmic-ray backgrounds, tighter cuts were applied.
These used the final $Z$-vertex
position, other tracking information and timing information.
Two additional cuts were also made~\cite{ZEUSdir+inc}:
\begin{itemize}
\item 
events with a well-identified positron candidate 
in the uranium calorimeter 
were removed.
\item 
a cut was made on the Jacquet-Blondel estimator of $y$~\cite{YJB},
$y_{JB} = \sum_i (E_i - E_{Z,i}) /2E_e$, where $E_{Z,i} = E_i
\cos\theta_i$, and $E_i$ is the energy deposited in the calorimeter
cell $i$ with a polar angle $\theta_i$ with respect to the
measured $Z$-vertex of the event.  The sum runs over all calorimeter
cells.  For any event where the scattered positron entered the uranium
calorimeter and was not well identified,
the value of $y_{JB}$ is close to one.
Proton beam-gas events have low values of $y_{JB}$.
To reduce further the contamination from this source,
it was required that $0.15 < y_{JB} < 0.7$.
This range corresponds approximately
to a true $y$ range of $0.2 < y < 0.8$, 
and so to a energy range 
of \mbox{$134 < W < 269$ GeV}
in the $\gamma p$ center-of-momentum frame.
\end{itemize}
These cuts restricted the range of $Q^2$ to be less than
$\sim 4$~GeV$^{2}$, with a median value
of about $10^{-3}$~GeV$^2$.

Dijet candidates were selected using the KTCLUS\cite{catani}
jet algorithm (details can be found in a previous 
ZEUS publication\cite{dijet_98}).
Cone\cite{snowmass} algorithms
were also used as a check: the conclusions did not change.
The jet
transverse energy measured in the ZEUS detector was corrected as a
function of pseudorapidity and transverse energy
to account for energy lost in inactive material. 
This correction was derived from the Monte Carlo (MC)
simulation described in the next section.
After all cuts, the kinematic region under study is defined by:
\mbox{$\ETJ>6$~GeV},
\mbox{$|\ETAJ |<2$},
\mbox{$Q^2<4$~GeV$^2$} and 
\mbox{$0.2<y<0.8$}.

Events with a leading neutron were selected from the inclusive dijet sample
by requiring a large energy deposit ($>400$ GeV) in the FNC.
The segmentation of the FNC
permits the identification of protons, photons, and neutrons. 
Scattered protons are bent into the top towers (11-14)
by the HERA dipole magnets.
As seen in Fig.~\ref{fig:fnc}, the geometric
aperture of the FNC for neutral particles at normal incidence 
is centered on towers 7 and 8. 
Scattered protons were eliminated from the sample
by requiring that the tower with the maximum energy deposit
be either 6,7,8 or 9. Although both photons and neutrons
produce large energy deposits in the bottom section
(towers 1-10), the vertical spread of electromagnetic showers is
much less than that of hadronic showers.
Photons were removed by eliminating showers with a small vertical spread
($\le 3$ cm).    
Finally, neutrons that started showering in front of the FNC were
removed by requiring that the scintillator veto counter farthest
from the FNC had an energy deposit below that of a minimum-ionizing particle.
Only the farthest counter was used, to minimize false vetoes due to
calorimeter albedo. Showers with spreads greater than 7.5~cm were
also removed since they are inconsistent with 
originating from a single high-energy hadron.  

After these cuts, 1921 events with a neutron remained, 
comprising approximately 1\% (before correction for the FNC acceptance) 
of the inclusive dijet sample of $2\cdot 10^5$ events.  


\section{Monte Carlo simulation}
\label{sec:montecarlo}

Monte Carlo simulations were used to correct the data for
acceptance and for smearing of the measured quantities due to the
finite resolution of the detector. 
For all generated events, the
ZEUS detector response was simulated in detail
using a program based on {\sc GEANT3.13}~\cite{geant},
and the Monte Carlo
events were subjected to the same analysis chain as the data. 
For the inclusive dijet analysis, the data were compared to Monte Carlo
simulations based on {\sc PYTHIA5.7}~\cite{PYTHIA} and
{\sc HERWIG5.9}~\cite{HERWIG}, which include leading-order QCD 
calculations. 
A minimum cut-off value
$\hat{p}_{\mbox{\rm\tiny T}}^{\mbox{\rm\tiny min}}$ of 2.5 GeV 
was applied at the MC generator level to
the transverse momenta of the outgoing partons
in the hard scattering process. 
The  HERWIG event generator was used to check
the PYTHIA results. 
In PYTHIA, the photon flux is calculated using the
Weizs\"{a}cker-Williams approximation. The parton densities used
were GRV LO~\cite{GRV} for the photon and CTEQ4~LO~\cite{CTEQ} for the
proton. The hadronization in PYTHIA was performed using the LUND string model
as implemented in JETSET~\cite{JETSET}.
In HERWIG, the hadronization of partons is based on a cluster model. 
For comparison, the LAC1~\cite{LAC1} parameterization
for the photon and MRSA~LO~\cite{MRSA} for the proton
were also used. 

Previous studies have shown that
including a simulation of multiparton interactions (MI)
in the parton shower programs significantly improves
the description of jet production in the 
forward region\cite{dijet_98}.
This option, which applies only to resolved processes, adds
interactions between the partons in the remnants of the proton 
and photon, calculated as LO QCD processes, 
to the hardest scattering process of the event.
It was implemented in the HERWIG simulation\cite{butterworth}.

The energy corrections for jets were determined
from the Monte Carlo samples by comparing the true transverse
energy of a jet (found by applying the algorithm to the final state
particles) to the transverse energy measured in the
calorimeter simulation.
The correction to the jet energies was on average
$+17\%$, varying
between $+10\%$ and $+25\%$ depending upon $\ETAJ $.
The largest corrections occurred
at boundaries within the calorimeter.
No correction was applied to the jet pseudorapidity, since
the average shift in $\eta$ between the true and detected jets
was less than $\pm$0.05 for all $\eta$ values in the range used for the
cross section measurements. In each event, the two jets with 
the highest transverse energies were selected.

For the dijet events with a neutron tag \mbox{($E_n>400$ GeV)}, 
the data were corrected using Monte Carlo 
programs based
on {\sc POMPYT2.5}~\cite{POMPYT} and {\sc RAPGAP2.06}~\cite{RAPGAP}.
These simulations include pion-exchange processes
where a virtual pion is emitted from the
incoming proton (see Appendix).
As discussed later, such a model reproduces the
characteristics of the data.
The POMPYT generator makes use of the program PYTHIA to
simulate $e^+\pi^+$ interactions via resolved and direct photon
processes. These programs simulate higher order effects
through the use of leading-order parton showers.
Hadronization processes are
implemented by JETSET (the LUND string model). 
In both programs GRV-LO was used 
for the parton densities of the photon and SMRS-P3~\cite{SMRS}
for the parton densities of the pion. 

The conclusions are independent of
the Monte Carlo model used for the corrections:
the jet energy
corrections and acceptance calculations, both inclusive and 
neutron-tagged,
can be performed with any of the four Monte Carlo
programs (POMPYT, RAPGAP, HERWIG, PYTHIA)
without significant change in the results. 


\section{Systematic uncertainties}

A detailed study of the sources contributing to the systematic
uncertainties of the measurements was carried out\cite{mohsen}.
Those
associated with the CTD and CAL, which impact on the
jet measurement, and those associated with
the FNC, which impact on the neutron measurement,
are considered separately.

For the jet measurements, 
the uncertainties are grouped into the following 
classes:
\begin{itemize}
  \item absolute energy scale of the CAL: \normalsize 
        the energy scale uncertainty of the low $E_T$ CAL 
        jets used in this study is 5\%, leading to 
        an uncertainty of 15 to 20\% on the cross section;
  \item model dependence: \normalsize for the inclusive
        dijet cross section, the jet-acceptance
        correction was performed using HERWIG instead of
        PYTHIA. For the neutron-tagged sample, the acceptance
        correction was changed from POMPYT to RAPGAP.
        The associated uncertainties were at the 5\% level; 
  \item parton parameterization: \normalsize changing the
        parton densities of the proton (CTEQ to MRSA) contributed
        an uncertainty of
        0.4\%; changing the parton densities of the photon (GRV to LAC1) 2\%;
  \item event selection: variation of 
        each selection cut by one standard deviation of the resolution
        gave uncertainties of about 5\%.
\end{itemize}

The main systematic uncertainties associated with the FNC were:
\begin{itemize}
  \item absolute energy scale of the FNC: the uncertainty on the FNC 
        energy scale is $\pm 2$\%\cite{calor97}. 
        This introduced a 1.5\% normalization
        uncertainty on the cross section;

  \item beam-gas background: charge-exchange processes 
        for beam-gas interactions can
        produce a high-energy nucleon which might overlap with a 
        dijet event. The uncertainty in the correction 
        of 2\% is less than 1\%;

 
  \item event selection: veto-counter noise (from beam halo and calorimeter
        albedo) and veto-counter inefficiencies resulted in a 2.5\%
        uncertainty;

  \item angular distribution of neutrons: the acceptance of the FNC 
        was uncertain due to uncertainties in the angular distribution 
        of the neutrons. The uncertainty in the
        acceptance was estimated by using different parameterizations of the
        pion flux. To obtain a model-independent
        estimate, the angular distribution at fixed $\xl$ 
        was assumed to fall exponentially
        with $\pt^2$, $dN/d\pt^2\propto \exp (-b \pt^2)$, and the 
        acceptance and its systematic uncertainty were determined by
        varying the slope parameter $b$ as a function of $\xl$ 
        within the limits allowed
        by the data\cite{fnt_osaka}.
        The resulting acceptance uncertainty for neutrons with energy 
        $E_n > 400$ GeV was 5.5\%;
  
  \item sensitivity of the description of the beam-line:
        the calculated acceptance of the
        FNC depends on a complete and accurate description
        of the proton beam-line
        between the interaction point and the calorimeter.
        Both the amount of inactive material and the alignment must be known.
        The model was tested by comparing the fraction of
        neutrons in the FNC surviving the
        selection cuts, according to RAPGAP, to that observed. 
        The discrepancy between the expected and observed fraction
        gave a normalization uncertainty of 6\%. 
\end{itemize}

The dominant systematic uncertainty is that
associated with the
energy scale of the CAL jets. 
The statistical
errors and the systematic uncertainties were added in quadrature
and are shown as error bars in the figures. The systematic
uncertainty associated with the absolute energy scale is shown as a
shaded band. 
The systematic uncertainties from the FNC give a 9\% normalization
uncertainty on the neutron-tagged cross section. Since this uncertainty
does not affect the shape of any distribution, it is not included
in the figures.
In addition, a 
normalization uncertainty of 1.5\% from the luminosity determination
was not included; this is not relevant in the 
measurement of the neutron-tagged to inclusive event-rate ratios.


\section{Some results and comparisons}
\label{sec:models}

The usual simulation models for hard photoproduction processes such as
HERWIG or PYTHIA contain a fraction of events with a leading neutron, 
although it is not, a priori,
expected that such simulations will properly describe particle
production in the proton fragmentation region.
For leading neutrons produced with high longitudinal momentum
and low transverse momentum, particle-exchange models 
may be more appropriate. In this case the LO cross section
can be expressed as:
\begin{eqnarray}
 \sigma^{ep}_{res} &=& \left[ 
  \int\int dydQ^2 f_{\gamma/e}(y,Q^2) \int\int dx_{L} dt
  f_{\pi/p}(x_{L},t) \sum_{i,j} 
  \int d\xg f_{i/\gamma}(\xg,\mu^2)
  \int dx_{\pi} f_{j/\pi}(x_{\pi},\mu^2)\right.\nonumber\\
  & &\hspace{4cm}\left. \sum_{k,l} \int d\hat{p}^2_{\mbox{\rm\tiny T}}
  \frac{d\hat{\sigma}_{i+j \rightarrow k + l }}{
  d\hat{p}^2_{\mbox{\rm\tiny T}}}(\hat{s},\hat{p}^2_{\mbox{\rm\tiny T}},\mu^2) \right]
\label{eqn:fac}
\end{eqnarray}
where the exchanged meson is denoted by $\pi$, and a sum
over all exchanged mesons is implied. 
The dijet cross section in charge-exchange photoproduction contains
contributions from both the direct and resolved processes.
In Eq.~(\ref{eqn:fac}), 
$f_{\gamma/e}$ is the splitting function of a positron
into a photon and positron,
$f_{\pi/p}$ is the splitting function of a proton into
a meson and neutron ( i.e., the flux of mesons in the proton),
$f_{i/\gamma}$ is the density of partons of type $i$ in the photon, 
and $f_{j/\pi}$ is the density of partons of type $j$ in the meson.
The sum in $i,j$ runs over 
all possible types of partons $i$ present in the photon 
and $j$ in the meson. The sum in $k,l$
runs over all possible types of final state partons.
The term $\hat{\sigma}_{i+j \rightarrow k+l}$ is the cross section for the
two-body collision $i + j \rightarrow k + l$ and depends on
the square of the center-of-momentum-frame energy,
$\hat{s}=y(1-\xl)\xg x_{\pi}s$, 
the transverse momentum of the two outgoing 
partons ($\hat{p}^2_{\mbox{\rm\tiny T}}$),
and the momentum scale ($\mu$) at which the strong-coupling 
constant ($\alpha_s(\mu^2)$)
is evaluated. For the direct process, Eq.~(\ref{eqn:fac}) also
holds except that $f_{i/\gamma}(x_{\gamma})$ is replaced by the 
Dirac delta function, $\delta(1-x_{\gamma})$, and there is no
sum over partons $i$ in the photon.

Equation~(\ref{eqn:fac}) 
incorporates the assumption of factorization.
In particular, the
resolved cross section depends on four ``parton'' densities and a
two-body scattering cross section. 
The kinematic variables $y$, $\xl$, $\xg$ and $x_{\pi}$ are coupled
only through the $\hat{s}$ dependence of the hard-scattering cross
section, $\hat{\sigma}$. 
A priori, the shape of the
neutron-tagged jet cross sections depends on the
kinematic variables of the neutron. 
In a complete factorization of the baryon and the photon vertices, 
however,
the shape of $\xg$ and the jet-variable distributions 
would not depend either on the presence of a neutron 
or explicitly on its kinematic variables. 
Similarly,
the energy spectrum of the neutrons would be independent of the
photon and jet variables.





\begin{figure}[htb]
\includegraphics[width=\linewidth]{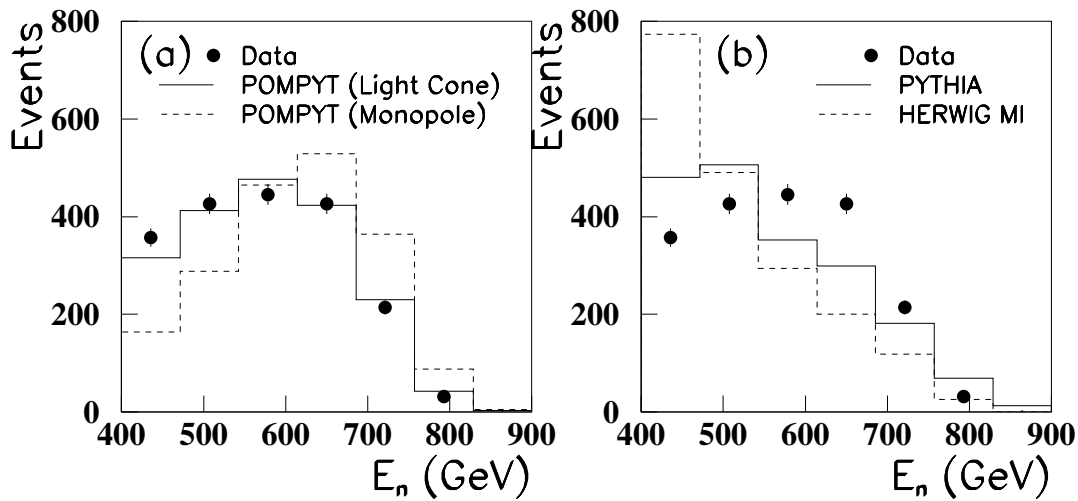}
\vspace{-8cm}
\caption{
Comparison of the neutron energy spectrum 
in dijet events with 
(a)
the predictions of POMPYT for 
(i) the light-cone and
(ii) the monopole 
pion form-factors and flux without the pion's Regge trajectory,
and (b) the predictions of 
PYTHIA and HERWIG with multiparton interactions.
The results for RAPGAP are indistinguishable from those of POMPYT.
The energy spectrum is uncorrected for acceptance.
The Monte Carlo results,
which take into account the acceptance and resolution of the FNC,
are area-normalized to the data.
Only statistical errors are shown.
}
\label{fig:fig4}
\end{figure}

Meson-exchange models, in the context of Regge theory, 
are often used to describe nucleon charge-exchange reactions.
Since the mass of the pion is
small compared to all other mesons, 
pion exchange (see Appendix)
is expected to dominate the 
amplitude for the $p\rightarrow n$ transition,
with small contributions from $\rho$, $a_2$, etc. 
In fact, LO one-pion-exchange models
account for both the shape and normalization
of the distributions for the neutron-tagged data. 
Figure~\ref{fig:fig4}(a)
compares the shape of the 
uncorrected
neutron energy spectrum to the prediction of the
one-pion-exchange (OPE) model POMPYT for the monopole and
light-cone pion form-factors.
The POMPYT OPE model with the light-cone form-factor
agrees well with the measured energy spectrum. 
RAPGAP gives essentially the same prediction.
The figure also shows
that the monopole choice for the form factor is 
disfavored.
It gives a distribution which is
both shifted in energy and too narrow. 
For this comparison, the \mbox{SMRS-P3~\cite{SMRS}}
pion parton densities have been used;
however, the predicted
shape of the neutron spectrum is
insensitive to the choice of parton
density of the pion\cite{SMRS,owen_pi,abfkw_pi,grv_pi} (not shown). 
These parton densities are constrained by 
dimuon and prompt-photon production
data from fixed-target experiments that
are sensitive mainly to the valence quark distributions,
and the parameterizations are similar
in the $x_{\pi}$ range studied here.

Figure~\ref{fig:fig4}(b)
shows a comparison of the shape of the neutron energy
distribution with the predictions of
the Monte Carlo programs PYTHIA and HERWIG.
Neither simulation provides a good description
of the data.
Moreover,
PYTHIA (HERWIG) predicts a leading neutron ($E_n > 400$ GeV)
in 2\% (0.5\%)
of the dijet events in comparison with the $4.9\pm 0.4$\% 
(after correction for the acceptance of the FNC) observed 
in the data. 

The Monte Carlo models considered here
do not contain an explicit diffractive 
component; 
however, neutrons can also be produced by 
the diffractive dissociation of the incoming proton
through Pomeron exchange. 
Monte Carlo studies indicate that
such neutrons will have an
energy spectrum which agrees qualitatively with 
that observed in the data. 
Diffractive processes give rise to a large rapidity gap between
the hadronic system and the remnant of the proton, which is either
a single proton or a low-mass system with the quantum  
numbers of the proton. 
The pseudorapidity of the most-forward
hadron 
($\eta_{\mbox{\rm\tiny max}}$)
in the central detector was used 
to select diffractive events.
Meson exchange can also give rise to events with a large
rapidity gap via
the Deck mechanism\cite{deck} in which  
the exchanged meson itself scatters diffractively
off the incoming photon and escapes undetected down
the beam pipe. 

\begin{figure}[htb]
\vspace{9pt}
\includegraphics[width=\linewidth]{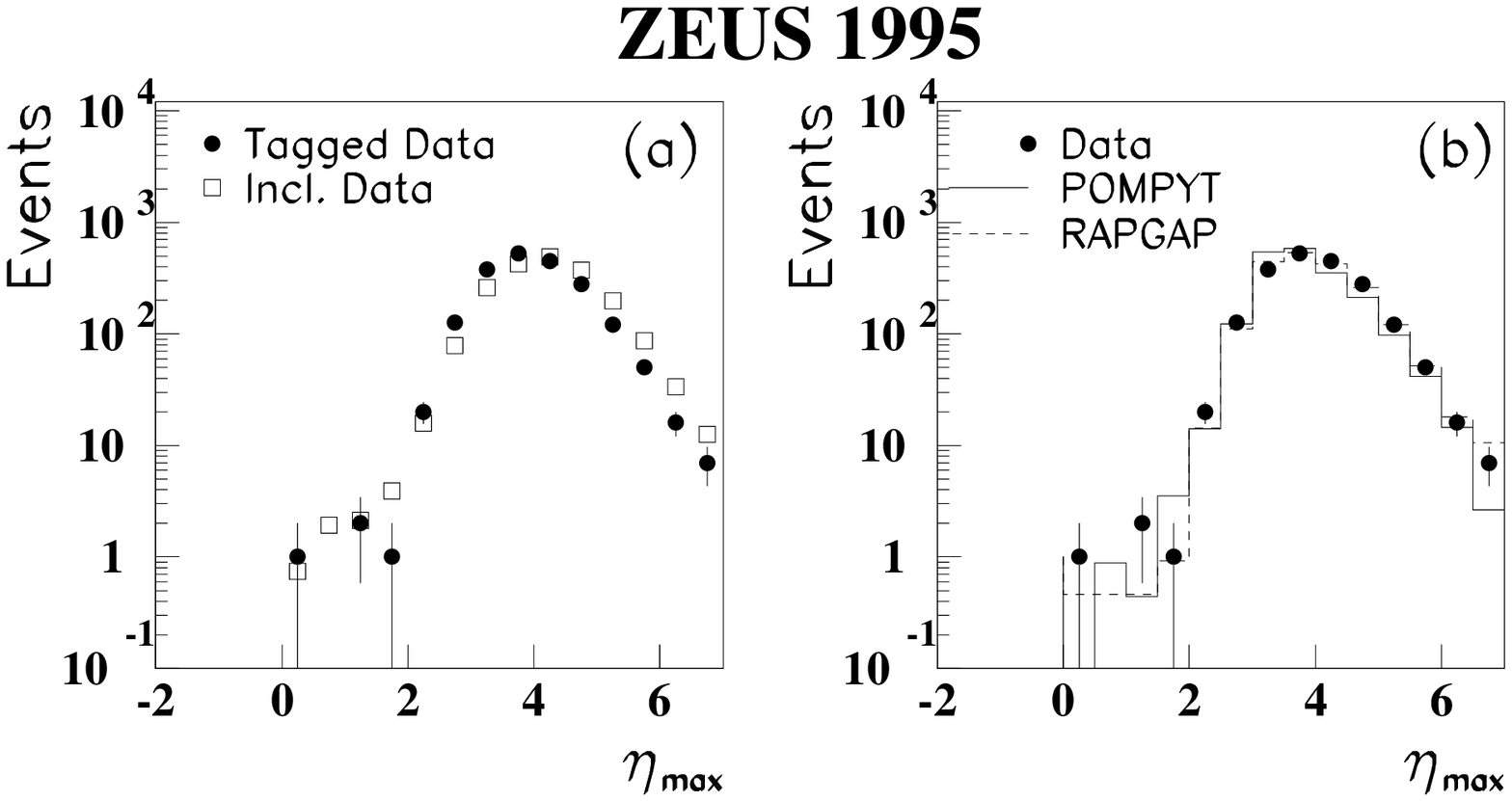}
\vspace{-8.cm}
\caption{
(a)
The $\eta_{\mbox{\rm\tiny max}}$ distribution for neutron-tagged 
and inclusive dijet photoproduction data. 
(b) 
The $\eta_{\mbox{\rm\tiny max}}$ distribution for neutron-tagged 
dijet photoproduction compared to the predictions
of POMPYT and RAPGAP.
The inclusive distribution is area normalized to the tagged
distribution.
Only statistical errors are shown.
}
\label{fig:figeta}
\end{figure}

Diffractively dissociating 
protons are expected to account for
only a small fraction of the dijet events,
both inclusive and neutron-tagged, 
because the $E_T$ of diffractive events is 
severely limited by the small fraction
of the proton energy carried by the Pomeron ($< 5$\%).
In contrast, in neutron production there is on average
a much larger
fraction of the initial proton energy ($\approx 25$\%)
available for jet production.
Figure~\ref{fig:figeta}(a)
shows the $\eta_{\mbox{\rm\tiny max}}$ distribution
for both inclusive and neutron-tagged dijet photoproduction,
normalized to equal area.
The shapes of the two distributions are similar.
Although there are differences between the two
distributions at large $\eta_{\mbox{\rm\tiny max}}$,
this region is insensitive to diffractive processes. 
Less than 1\% of each sample
has a large rapidity gap ($\eta_{\mbox{\rm\tiny max}}<2$).
In addition, the distributions of the jet variables, $\ETJ$,
$\ETAJ $, and $\xgo$ for events with a large
rapidity gap\cite{zeus_dijet_diff} are different from those of the 
neutron-tagged sample, which strongly resemble the
inclusive sample ( see Section~\ref{sec:factorization}).
POMPYT and RAPGAP also reproduce the $\eta_{\mbox{\rm\tiny max}}$ 
distribution of neutron-tagged events, 
as seen in Fig.~\ref{fig:figeta}(b), although neither
Monte Carlo contains an explicit diffractive
component.

Neutrons can also be produced indirectly through the production
and decay of baryonic resonances. Most prominent of these
is the $\Delta$, 
which is itself produced directly through
$\pi$, $\rho$ or $a_2$ exchange, and which can decay via
\mbox{$\Delta^0\rightarrow n\pi^0$} or 
\mbox{$\Delta^+\rightarrow n\pi^+$}.
Monte Carlo studies indicate that
such decay neutrons will also have an
energy spectrum which agrees qualitatively with the data; 
however,
the study of hadronic interactions\cite{erwin_delta,higgins,barish,dao}
shows that 
the $\Delta\pi$ contribution to the Fock state of the proton
is approximately half that of $n\pi$\cite{holtmann,thomas}.
When account is taken of the branching ratio for
$\Delta\rightarrow n$, 
such indirect neutron production is
small compared to
the direct $p\rightarrow n$ transition.

It can therefore be concluded that
standard fragmentation processes,
diffractively dissociating protons, and 
the decay of the $\Delta$ resonance
are ruled out as the dominant source of
dijet events tagged with a leading neutron.
The data will be further compared with the
one-pion-exchange model in Section~\ref{sec:ope}.



\section{Factorization tests}
\label{sec:factorization}

\begin{figure}[htb]
\includegraphics[width=\linewidth]{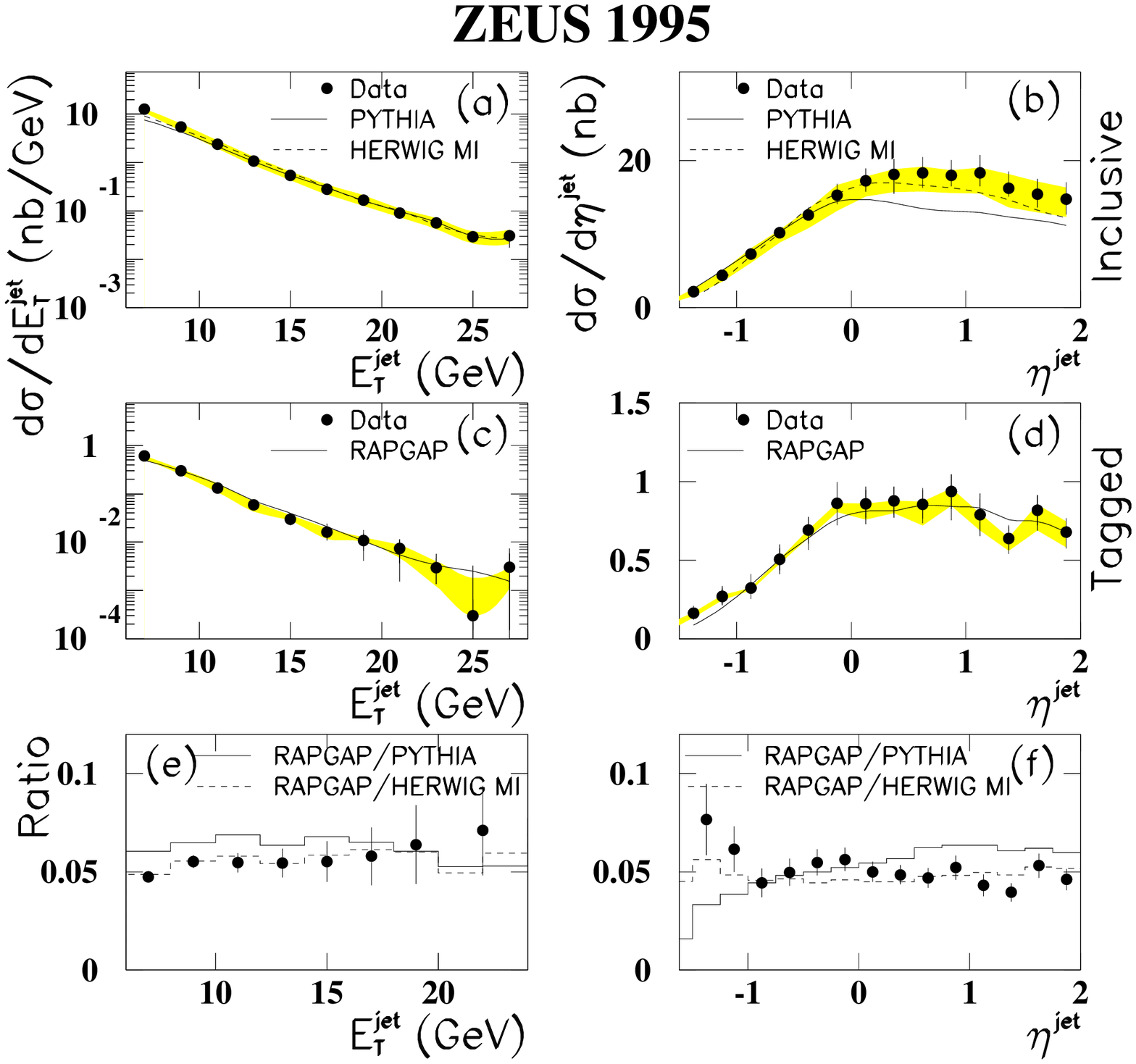}
\vspace{-2.cm}
\caption{The differential dijet cross sections as a function of
$\ETJ $ and $\ETAJ $ for both inclusive (a,b) and
neutron-tagged (c,d) photoproduction.
The kinematic region studied is
\mbox{$\ETJ>6$~GeV},
\mbox{$|\ETAJ |<2$},
\mbox{$Q^2<4$~GeV$^2$},  
\mbox{$0.2<y<0.8$},
\mbox{$E_n>400$ GeV}, and
\mbox{$\theta_n < 0.8$ mrad}.
For the inclusive cross sections the Monte Carlo
predictions have been normalized to the measured cross sections at
high $\ETJ $ and negative $\ETAJ $, respectively,
in order to facilitate the
comparison of the shapes. 
The tagged data are compared with the predictions of the 
one-pion-exchange model RAPGAP using the SMRS-P3
pion parton densities and the light-cone pion form-factor.
The bands show the systematic uncertainty due to the calorimeter energy
scale. 
The error bars show the statistical error added in quadrature
with the remaining systematic error.
The tagged cross sections have an additional normalization
uncertainty of 9\% which is not shown.
Also shown are the ratio of the cross sections 
of neutron-tagged to inclusive dijet photoproduction 
as a function of 
(e) $\ETJ$, i.e. $\left[\mathrm{(c)/(a)}\right]$, 
and (f) $\ETAJ $, i.e. $\left[\mathrm{(d)/(b)}\right]$;
and Monte Carlo predictions for the ratio using
RAPGAP to PYTHIA and RAPGAP to HERWIG.
}
\label{fig:fig6}
\end{figure}

The differential 
dijet cross sections as a function of $\ETJ$
and $\ETAJ $ are shown in Fig.~\ref{fig:fig6}(a,b)
for the inclusive sample, and in Fig.~\ref{fig:fig6}(c,d)
for the neutron-tagged sample.
The predictions of PYTHIA and HERWIG
describe 
the shape of the $\ETJ$ distribution
reasonably well.
The predictions
have been normalized to the measured cross section at high
\mbox{$\ETJ$} \mbox{($>$16 GeV)} 
in order to facilitate the
shape comparison. 
For the $\ETAJ $ distribution,
the Monte Carlo is normalized at small $\ETAJ $, 
\mbox{$-1.5<\ETAJ <0$}.
HERWIG with multiparton interactions 
is in better agreement with the data than PYTHIA without 
such interactions.

The study of ratios of neutron-tagged to inclusive
cross sections is advantageous since 
many systematic uncertainties, 
especially those related to the jet measurements,
are greatly reduced. 
In addition, the ratio provides a quantitative 
comparison of the shapes of the cross sections,
and so tests factorization.
That the  neutron-tagged and inclusive differential cross
sections have similar shapes as a function of $\ETJ$
is evident in
Fig.~\ref{fig:fig6}(e).
The cross sections fall
by over two orders of magnitude in the range $6 < \ETJ <25$ GeV,
but the ratio is approximately constant
as a function of $\ETJ$.
Figure~\ref{fig:fig6}(f) shows that the ratio as a function of
$\ETAJ $ falls slightly with increasing $\ETAJ$. 
The ratios of RAPGAP to PYTHIA and 
RAPGAP to HERWIG with multiparton interactions are 
also shown in Figs.~\ref{fig:fig6}(e,f). The ratio of RAPGAP to
HERWIG is in better agreement with the data, as expected from the
cross section comparisons.
The agreement of RAPGAP and POMPYT with the tagged data supports 
factorization, which is built into these MC models.

\begin{figure}[htb]
\vspace{9pt}
\includegraphics[width=\linewidth]{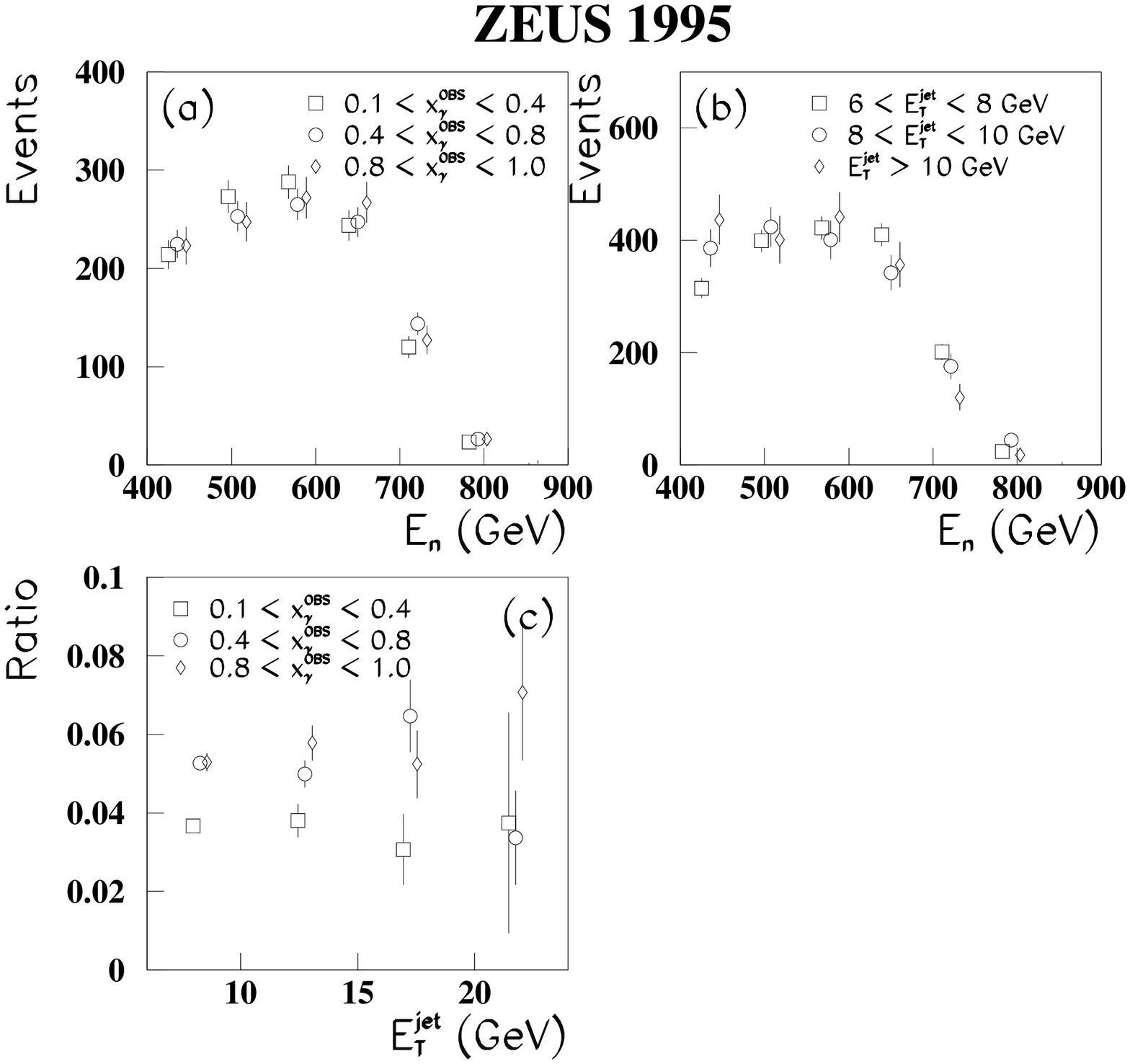}
\vspace{-2.cm}
\caption{
Uncorrected neutron energy spectra for different ranges 
of (a) $\xgo$ and (b) $\ETJ $. 
The spectra are area-normalized. 
(c) The ratio of the cross-sections of neutron-tagged to 
inclusive dijet photoproduction as a function of 
$\ETJ$
for different ranges of $\xgo$
(applied to both the inclusive and tagged samples). 
The points are offset slightly to improve visibility.
Only statistical errors are shown. There is an overall
normalization uncertainty (not shown) of 9\% for (c).}
\label{fig:fig7}
\end{figure}

To test further the factorization properties 
of the dijet cross section,
the neutron-tagged and inclusive samples were divided into bins of 
$\xgo$ and $\ETJ$. 
The shape of the observed energy
spectrum of the neutron is approximately independent 
of $\xgo$ and $\ETJ $,
as seen in Fig.~\ref{fig:fig7}(a) and (b).
Moreover, in a given bin of $\xgo$,
the fraction of events with a leading neutron
is approximately independent on $\ETJ$, 
as seen in Fig.~\ref{fig:fig7}(c). 

\begin{figure}[htb]
\includegraphics[width=\linewidth]{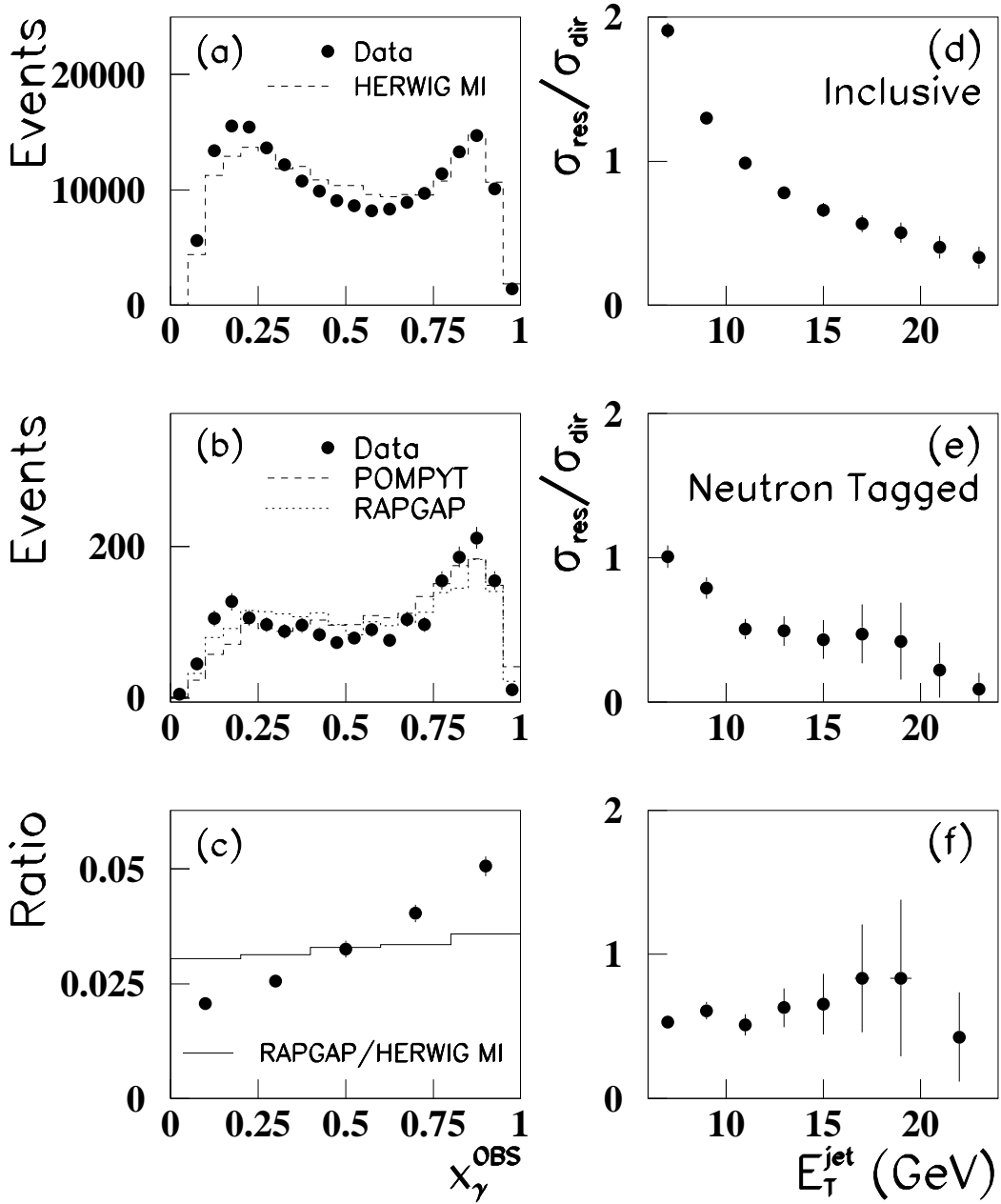}
\vspace{-1cm}
\caption{
The uncorrected $\xgo$ distribution for (a) 
inclusive events
compared to the expectations of HERWIG with multiple
interactions, and (b) for neutron-tagged events
compared to the expectations of POMPYT and RAPGAP. 
The ratio
of the neutron-tagged to inclusive $\xgo$ distributions is
shown in (c) together with the ratio of RAPGAP to HERWIG MI.
The Monte Carlo predictions are area normalized to the data
in (a,b) and normalized at $\xgo=0.5$ in (c).
The ratio of the resolved to 
direct cross sections as a function of $\ETJ$
is shown in (d) and (e) 
for inclusive  and neutron-tagged photoproduction,
respectively.
Resolved (direct events) are defined by 
$\xgo <0.75$ ($\xgo >0.75$).
The ratio of the ratios shown in (d) and (e),
i.e. $\left[\mathrm{(e)/(d)}\right]$,
is shown in (f).
Only statistical errors are shown. 
}
\label{fig:fig8}
\end{figure}

The $\xgo$ distribution is
determined by the parton densities in the colliding particles,
kinematic factors, and a possible presence of
interactions between the hadron remnants.
Figures~\ref{fig:fig8}(a,b)
show the uncorrected $\xgo$ distribution for
inclusive and neutron-tagged events, respectively. 
The corresponding
ratio of neutron-tagged to inclusive production is
shown in Fig.~\ref{fig:fig8}(c), corrected only for the
acceptance of the FNC. 
In sharp contrast to the results shown in Fig.~\ref{fig:fig6}, 
the ratio rises with increasing $\xgo$. The rise is only
partially explained by the Monte Carlo models.
The Monte Carlo predictions are shown area normalized to 
the data in Figs.~\ref{fig:fig8}(a,b); 
the ratio of RAPGAP to HERWIG normalized 
at $\xgo=0.5$ is shown in Fig.~\ref{fig:fig8}(c).
 
Figure~\ref{fig:fig8}(d) shows the ratio of
the resolved ($\xgo <0.75$)
and direct ($\xgo >0.75$) photoproduction 
cross sections as
a function of $\ETJ$ for inclusive events while
Fig.~\ref{fig:fig8}(e) show the same quantity for
neutron-tagged events. The size of the direct
component increases with $E^{\mbox{\rm\tiny jet}}$ 
for both the neutron-tagged and
the inclusive samples.
Overall the direct component is approximately
twice as large in the neutron-tagged sample. 
Figure~\ref{fig:fig8}(f) shows that
the ratio of the two ratios 
is constant within errors as a function 
of $\ETJ$ indicating that, 
although the proportion changes,
the shape in $\ETJ$ is the same
for both samples.

In summary, the shape of the neutron energy 
spectrum is approximately independent of the the jet variables
$\ETJ$, $\ETAJ$ and $\xgo$;
the jet variables $\ETJ$ and $\ETAJ$ are
relatively insensitive to the presence of a neutron;
however, the ratio of neutron-tagged to inclusive dijet
events rates rises with increasing $\xgo$. 

\section{Comparison to a one-pion-exchange model}
\label{sec:ope}

In the previous two sections, it was shown
that
standard fragmentation processes,
diffractively dissociating protons, and 
the decay of the $\Delta$ resonance
are not the dominant source of
dijet events tagged with a leading neutron,
but that the data are  consistent with 
one-pion exchange and approximately
satisfy factorization. 
It is therefore appropriate in this section to make
further comparisons to OPE
under the assumption that it is the dominant
mechanism for the production of the 
neutron-tagged events.

The predictions of RAPGAP simulations
using the SMRS-P3
parton densities for the pion and the light-cone
pion form-factor 
are in good agreement with the $\ETJ$ spectrum for the
neutron-tagged events of Fig.~\ref{fig:fig6}(c).
POMPYT, which is not shown, gives a similar result.
Both OPE Monte Carlo models also adequately describe 
the shape of the $\ETAJ $
distribution of Fig.~\ref{fig:fig6}(d), where only
RAPGAP is shown.
The leading-order OPE model is also able to
account for the normalization of the data,
as shown by the RAPGAP prediction.
In contrast to the inclusive case (Fig.~\ref{fig:fig6}(b)), 
the Monte Carlo simulations reproduce the forward $\ETAJ $ region
without a simulation of multiparton interactions. 
This comparison suggests that
hadron remnant interactions are 
less important for the neutron-tagged sample,
in agreement with naive expectations based
on their lower effective center-of-momentum-frame energy
compared to the inclusive sample.
The RAPGAP and POMPYT simulations are also in fair agreement
with the $\xgo$ distribution
for the neutron-tagged events (Fig.~\ref{fig:fig8}(b)).
As in the case of the $\ETAJ$ distribution, the OPE 
Monte Carlo models describe the data without simulation
of MI, which is needed in the case of the inclusive
sample (Fig.~\ref{fig:fig8}(a)). 
According to the Monte Carlo models, the MI contribution
in the inclusive sample strongly increases at low values
of $\xgo$\cite{dijet_98}. Therefore the ratio of neutron-tagged
to inclusive events, which increases with $\xgo$ (Fig.~\ref{fig:fig8}(c)),
can be interpreted to be at least partially caused by the
difference in remnant-remnant interactions of the two samples.   

\begin{figure}[htb]
\includegraphics[width=\linewidth]{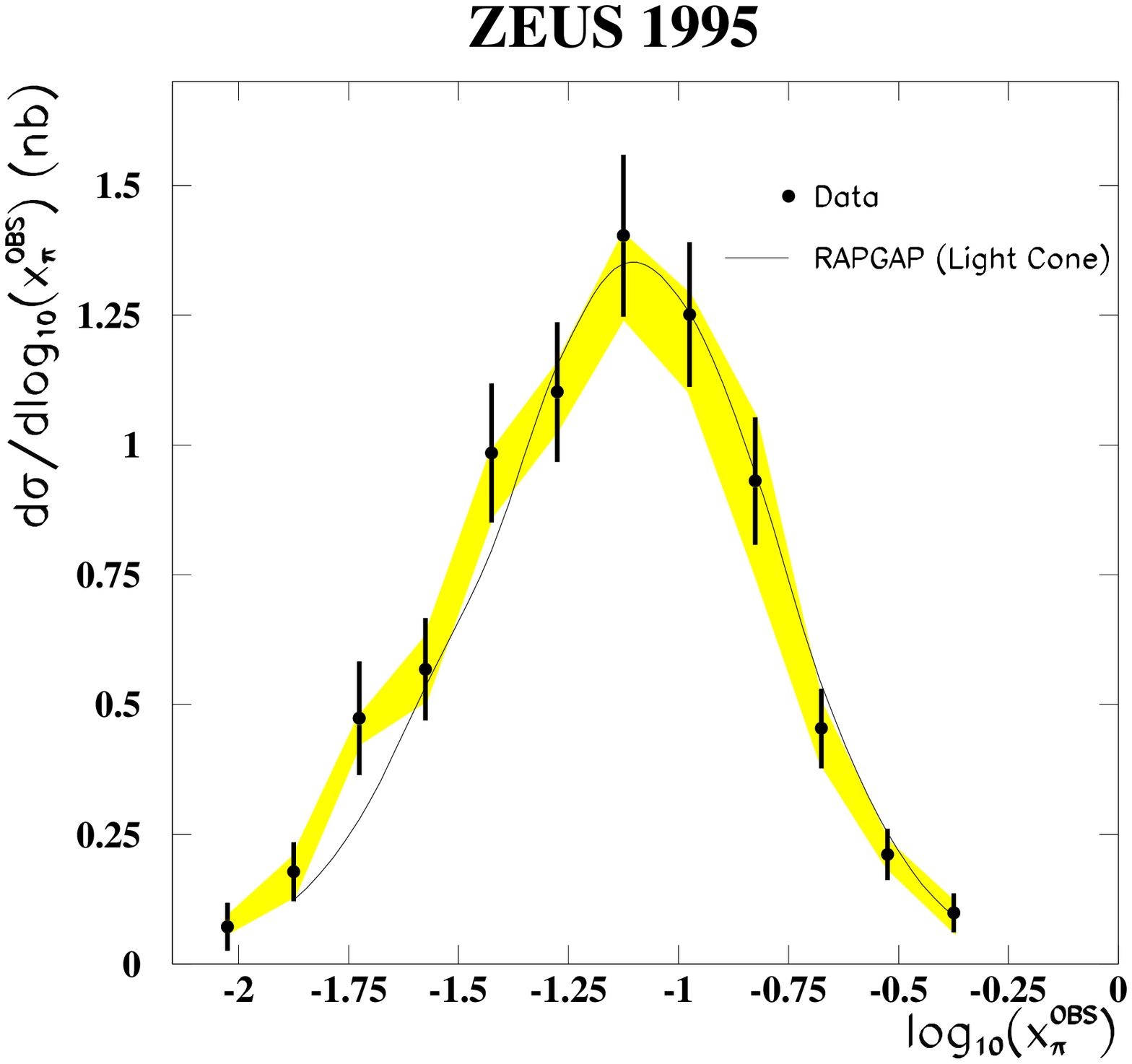}
\vspace{-2.cm} 
\caption{Differential cross section as a function of $\xpio$, the
fraction of the exchanged pion's momentum participating in the
production of the dijet system for the neutron-tagged sample. 
The measured cross section is compared to
the prediction of the Monte Carlo model RAPGAP
using the light-cone pion form-factor and
the SMRS-P3 pion parton densities.
The bands show the systematic uncertainty due to the calorimeter
energy scale. 
The error bars show the statistical error added in quadrature
with the remaining systematic error.
There is an additional normalization error of
9\% which is not shown.
}
\label{fig:tag_pion}
\end{figure}

The differential cross section for neutron-tagged events as a function of 
$\xpio$ (see Eq.~\ref{eqn:xpi}) 
is shown in Fig.~\ref{fig:tag_pion}. 
The measured cross section is compared
in Fig.~\ref{fig:tag_pion} to
the predictions of the RAPGAP Monte Carlo model
using the light-cone pion form-factor
and the SMRS-P3 pion parton densities.
The shape of the $\xpio$ distribution disfavors the
monopole form-factor (not shown), but both the shape
and normalization are reproduced by both RAPGAP and
POMPYT (not shown).
The results for
other parameterizations of the pion parton densities,
which are determined from hadron-hadron interactions,
are indistinguishable.
The bands show the systematic uncertainty due to the calorimeter
energy scale. There is an additional normalization error of
9\% which is not shown.

\section{Conclusions} 

Differential cross sections for the
inclusive reaction
\mbox{$e^+ p \rightarrow e^+ + \mathrm{jet} + \mathrm{jet} + X_r$}
and the neutron-tagged reaction
\mbox{$e^+ p \rightarrow e^+ + n + \mathrm{jet} + \mathrm{jet} + X_r$}
have been measured in photoproduction 
for $Q^2<4$ GeV$^2$ and $0.2<y<0.8$,
with jet transverse energy $\ETJ>6$ GeV, neutron
energy $E_n>400$ GeV and neutron production angle $\theta_n<$ 0.8 mrad.
Such neutrons are observed in $4.9\pm 0.4$\% of the events.

The shape of the neutron energy spectrum is observed to
be independent of both $\ETJ$
and $\xgo$. In addition, the ratio of neutron-tagged to inclusive
dijet production rates is approximately independent of the jet
transverse energy and pseudorapidity. These observations support the idea
of factorization of the positron and proton vertices.

The above ratio does depend on $\xgo$, the fraction of
the initial photon momentum participating in the hard interaction;
the direct to resolved fraction is approximately twice as large in the
neutron-tagged sample as in the inclusive sample.

The $\xgo$ distribution for the inclusive dijet process can be
reproduced in a simulation which includes multiparton interactions.
In contrast the neutron-tagged data are well simulated by LO Monte Carlo
models including one-pion exchange but without multiparton
interactions. These comparisons suggest that the rising ratio of
neutron-tagged to inclusive dijet rates as a function of $\xgo$ is at
least partially due to the difference in remnant-remnant interactions 
in the two samples.

The standard photoproduction Monte Carlo models, PYTHIA and HERWIG, fail to
reproduce either the neutron production rate or the neutron energy
spectrum. In contrast,
LO Monte Carlo models such as POMPYT and RAPGAP, which include 
one-pion exchange
and which assume factorization of the pion flux and pion structure,
reproduce all aspects of the neutron-tagged data for both the neutron
and jet kinematic variables. 

\section*{Acknowledgments}

We are especially grateful to the
DESY Directorate whose encouragement and financial support made
possible the construction and installation of the FNC.
We are also happy to acknowledge the efforts of
the DESY accelerator group and the support of the DESY computing group.
We thank H.\ Jung and G.\ Ingelman for their help with
the Monte Carlo programs RAPGAP and POMPYT.

\section*{Appendix: pion exchange}

Although there are many choices found in the literature
for the splitting function of a proton into a pion
and neutron,
they all can be summarized conveniently in the 
form\cite{bishari,field}:
\begin{eqnarray}
f_{\pi /p}(\xl ,t) = \frac {1}{4 \pi}\frac{g_{n \pi p}^2}{4 \pi}
\frac{-t}{(m^2_{\pi}-t)^2}(1-\xl )^{1-2 \alpha_ \pi (t)}
\left(F(\xl ,t)\right)^2
\end{eqnarray}
where $g_{n \pi p}$ is the coupling at the $n \pi p$ vertex, 
$m_{\pi}$ is the mass of the pion, and
$\alpha_{\pi}(t)=\alpha '(t-m^2_{\pi})$ is
the Regge trajectory of the pion. 
The trajectory is often omitted by
setting the slope constant $\alpha'= 0$, instead of
to the measured value, $\alpha'\simeq 1$ GeV$^{-2}$. 
$F(\xl ,t)$ is a form-factor which accounts 
for final state rescattering of the neutron.
Possible choices for the 
form-factor $F(\xl ,t)$ are:
\begin{equation}
 F(\xl ,t)=\left\{
  \begin{array}{ll}
  \exp\left(b(t-m^2_{\pi})\right)                                 &\mathrm{Exponential}\\
  \exp\left(R^2(t-m^2_{\pi})/(1-\xl )\right)                       &\mathrm{Light\ cone}\\
  \left(1-m^2_{\pi}/\Lambda^2\right)/\left(1-t/\Lambda^2\right)   &\mathrm{Monopole}
\end{array} \right.
\label{eqn:form}
\end{equation}
where $b$, $R$ and $\Lambda$ are constants\cite{holtmann,kopeliovich,frank89}. 
The choice of form-factor depends on whether
the pion's Regge trajectory is included or excluded.
For the flux with the Regge trajectory the form factor is usually chosen 
to be an exponential\cite{bishari}. 
The light-cone form-factor is usually associated with
the flux without the Regge trajectory\cite{zoller,pumplin}. 
These choices are associated with the
experimental observation that in hadronic interactions the shape of 
the $\xl $ distribution depends on $t$\cite{engler,robinson,hartner}. 

The experimental data are consistent 
with the flux with the trajectory term included 
and a constant form factor 
(that is, $b\approx 0$)\cite{erwin,pickup,bishari,field,engler,robinson,flauger,hanlon,hartner,eisenberg,blobel,abramowicz}.
Beam-gas data
at HERA are also consistent with
a small value for $b$\cite{calor97}; Kopeliovich 
et al\cite{kopeliovich} use $b=0.3$ GeV$^{-2}$. 
For the light-cone form-factor and the flux without the
pion's Regge trajectory, 
Holtmann et al.\cite{holtmann} take
$R=0.6$ GeV$^{-1}$. A calculation\cite{meissner}
of the pion form-factor for the one-boson-exchange potential (OBEP) 
suggests that the monopole with $\Lambda=0.8$ GeV is a good 
approximation; however,
RAPGAP (see Section~\ref{sec:montecarlo})
takes $\Lambda=0.5$ GeV\cite{frank89} and the
flux without the Regge trajectory. In the region of the
peak of the energy spectrum, $\xl \approx 0.7$, the Regge and 
light-cone splitting functions differ by $\lap 10$\%.   
It should be noted that choices other than those
given in Eq.~(\ref{eqn:form}) are possible. In particular,
recent studies\cite{holzwarth,fries} suggest that the
the experimental data is best described by a form factor that is
hard for small momentum transfers and soft for large momentum transfers.

Other isovector meson exchanges, 
such as the $\rho$ or $a_2$,
can also contribute
to direct neutron production. 
Recent theoretical studies of neutron production 
in $ep$ collisions 
show that processes other than direct pion exchange
are expected to contribute 
$\lsim 25$\% of neutron production\cite{holtmann,kopeliovich,thomas,nsss}.
These backgrounds to OPE, which increase the rate of neutron production
in the FNC phase space, are offset by absorptive rescattering of the 
neutron, which decreases the rate by approximately
the amount of the increase\cite{alesio,nszak}. 
Also absorptive rescattering preferentially removes neutrons with
larger $\pt$, increasing the pion contribution relative to the
$\rho$ and $a_2$. Therefore these effects are also neglected in the 
present analysis. 


\newpage

\clearpage

\clearpage

\clearpage

\clearpage

\clearpage

\clearpage

\clearpage

\clearpage

\begin{thebibliography}{999} 


\bibitem{erwin}
A.\ R.\ Erwin et al.,
\Journal{\PRL}{6}{1961}{628}.

\bibitem{pickup}
E.\ Pickup, D.\ K.\ Robinson and E.\ O.\ Salant,
\Journal{\PRL}{7}{1961}{635}.

\bibitem{engler}
J.\ Engler et al., \Journal{\NPB}{84}{1975}{70}.

\bibitem{robinson}
B.\ Robinson et al., \Journal{\PRL}{34}{1975}{1475}.

\bibitem{flauger}
W.\ Flauger and F.\ M\"onnig, \Journal{\NPB}{109}{1976}{347}.

\bibitem{hanlon}
J.\ Hanlon et al., \Journal{\PRL}{37}{1976}{967};
\Journal{\PRD}{20}{1979}{2135}.

\bibitem{hartner}
G.\ Hartner, 
Ph.\ D.\ Thesis, McGill University, 1977 (unpublished).

\bibitem{eisenberg}
Y.\ Eisenberg et al., \Journal{\NPB}{135}{1978}{189}.

\bibitem{blobel}
V.\ Blobel et al., \Journal{\NPB}{135}{1978}{379}.

\bibitem{abramowicz}
H.\ Abramowicz et al., \Journal{\NPB}{166}{1980}{62}.

\bibitem{yukawa}
H.\ Yukawa, \Journal{Proc.\ Phys.\ Math.\ Soc.\ Japan\ }{17}{1935}{48}. 

\bibitem{sullivan}
J.\ D.\ Sullivan, \Journal{\PRD}{5}{1972}{1732}.

\bibitem{bishari}
M.\ Bishari, \Journal{\PLB}{38}{1972}{510}.

\bibitem{field}
R.\ D.\ Field and G.\ C.\ Fox, \Journal{\NPB}{80}{1974}{367}.

\bibitem{ganguli}
S.\ N.\ Ganguli and D.\ P. Roy, \Journal{\PRep}{67}{1980}{201}.

\bibitem{zakharov}
B.\ G.\ Zakharov and V.\ N.\ Sergeev, \Journal{\SNP}{38}{1983}{947};
\Journal{\SNP}{39}{1984}{448}.

\bibitem{zoller}
V.\ R.\ Zoller, \Journal{\ZPC}{53}{1992}{443}.

\bibitem{fnc2}
ZEUS Collaboration, M.\ Derrick et al.,
\Journal{\PLB}{384}{1996}{388}.

\bibitem{h1f2lb}
H1 Collaboration, C.\ Adloff et al., \Journal{\EPC}{6}{1999}{587}.

\bibitem{zeusdij95}
ZEUS Collaboration, M.\ Derrick et al., \Journal{\PLB}{348}{1995}{665}.

\bibitem{dijet_98}
ZEUS Collaboration, J. Breitweg et al., \Journal{\EPC}{1}{1998}{109}.

\bibitem{zeus_dijet_diff}
ZEUS Collaboration, J.\ Breitweg et al., \Journal{\EPC}{5}{1998}{41}.

\bibitem{owens}
J.\ F.\ Owens, \Journal{\PRD}{21}{1980}{54}.

\bibitem{drees}
M.\ Drees and F.\ Halzen, \Journal{\PRL}{61}{1988}{275};\\
\newblock M.\ Drees and R.\ M.\ Godbole, \Journal{\PRD}{39}{1989}{169}.

\bibitem{zeusnov93}
ZEUS Collaboration, M.\ Derrick et al., \Journal{\PLB}{322}{1994}{287}.

\bibitem{zeusdet} 
ZEUS Collaboration, M.\ Derrick et al., \Journal{\PLB}{293}{1992}{465}; \\
\newblock ZEUS Collaboration, `The ZEUS Detector',
Status Report 1993, ed. U.\ Holm,  
DESY(1993) (unpublished).

\bibitem{ctd}  
N. Harnew et al., \Journal{\NIMA}{279}{1989}{290};\\
\newblock B. Foster et al., \Journal{\NPPS}{32}{1993}{181};
\Journal{\NIMA}{338}{1994}{254}.

\bibitem{cal}  
M.\ Derrick et al.,  \Journal{\NIMA}{309}{1991}{77}; \\
\newblock A.\ Andresen et al., \Journal{\NIMA}{309}{1991}{101}; \\
\newblock A.\ Bernstein et al., \Journal{\NIMA}{336}{1993}{23}.

\bibitem{fnc3}
S.\ Bhadra et al., \Journal{\NIMA}{394}{1997}{121}.

\bibitem{fnt_osaka}
ZEUS Collaboration, 
Paper 440 submitted to 
XXXth International Conference on High Energy Physics, 
Osaka, Japan, 2000.

\bibitem{calor97}
ZEUS FNC Group, S. Bhadra et al., 
in {\it Proceedings of the Seventh
International Conference on 
calorimetry in High Energy Physics},
Tucson, Arizona, USA, November 9-14, 1997, 
eds. E. Cheu et al,
(World Scientific, Singapore, 1998).

\bibitem{fnc1}
S.\ Bhadra et al., \Journal{\NIMA}{355}{1995}{470}.

\bibitem{holtmann}
H.\ Holtmann et al., 
\Journal{\PLB}{338}{1994}{363}.

\bibitem{F2} 
ZEUS Collaboration, M. Derrick et al., \Journal{\PLB}{316}{1993}{412}.

\bibitem{ZEUSdir+inc} 
ZEUS Collaboration, M.\ Derrick et al., \Journal{\PLB}{322}{1994}{287}. 

\bibitem{YJB} 
F.\ Jacquet and A.\ Blondel, in
{\it Proceedings of the study of an ep facility for Europe}, 
ed.\ U.\ Amaldi, (DESY 79/48, Hamburg, 1979) p. 391.

\bibitem{catani}
S.\ Catani et al., 
\Journal{\NPB}{406}{1993}{187}.

\bibitem{snowmass}
J.\ Huth et al., in {\it Proceedings of the 1990 DPF Study on
High Energy Physics}, Snowmass, Colorado, ed. E.\ L. Berger
(World Scientific, Singapore, 1992) p. 134; \\
\newblock UA1 Collaboration,
G.\ Arnison et al., \Journal{\PLB}{123}{1983}{115}.

\bibitem{geant}  
GEANT 3.13, R. Brun et al., CERN DD/EE/84-1 (1987).

\bibitem{PYTHIA}
H.-U.\ Bengtsson and T.~Sj{\"o}strand, \Journal{\CPC}{46}{1987}{43}.

\bibitem{HERWIG}
G.\ Marchesini et al., \Journal{\CPC}{67}{1992}{465}.

\bibitem{GRV} 
M.\ Gl\"uck, E.\ Reya and A.\ Vogt, \Journal{\PRD}{46}{1992}{1973}.

\bibitem{CTEQ}
CTEQ Collaboration, J.\ Botts et al., \Journal{\PLB}{304}{1993}{159};\\
\newblock CTEQ Collaboration, H.\ L.\ Lai et al., \Journal{\PRD}{51}{1995}{4763}. 

\bibitem{JETSET} 
T.\ Sj\"ostrand, \Journal{\CPC}{82}{1994}{74}.

\bibitem{LAC1} 
H.\ Abramowicz, K.\ Charchula and A.\ Levy, \Journal{\PLB}{269}{1991}{458}.

\bibitem{MRSA} 
A.\ D.\ Martin, W.\ J.\ Stirling and R.\ G.\ Roberts, 
\Journal{\PRD}{50}{1994}{6734}.

\bibitem{butterworth}
J.\ M.\ Butterworth, J.\ R.\ Forshaw and M.\ H.\ Seymour,
\Journal{\ZPC}{72}{1996}{637}.

\bibitem{POMPYT}
P.\ Bruni and G.\ Ingelman, \Journal{\PLB}{311}{1993}{317};\\
\newblock P.\ Bruni, A.\ Edin and G.\ Ingelman, DESY 93-137 (1993), 
http://www3.tsl.uu.se/ thep/pompyt/.

\bibitem{RAPGAP}
H.\ Jung, \Journal{\CPC}{86}{1995}{147}.

\bibitem{SMRS}
P.\ J.\ Sutton et al., 
\Journal{\PRD}{45}{1992}{2349}.

\bibitem{mohsen}
M.\ Khakzad, 
Ph.\ D.\ Thesis, York University, 2000 (DESY-THESIS-2000-008).


\bibitem{owen_pi}
J.\ F.\ Owens, \Journal{\PRD}{30}{1984}{943}.

\bibitem{abfkw_pi}
P.\ Aurenche et al., 
\Journal{\PLB}{233}{1989}{517}.

\bibitem{grv_pi}
M.\ Gl\"uck, E.\ Reya and A.\ Vogt, \Journal{\ZPC}{53}{1992}{651}.

\bibitem{deck}
S.\ D.\ Drell and K.\ Hiida, \Journal{\PRL}{7}{1961}{199};\\
\newblock R.\ T.\ Deck, \Journal{\PRL}{13}{1964}{169}.

\bibitem{erwin_delta}
J.\ Erwin et al., \Journal{\PRL}{35}{1975}{980}.


\bibitem{higgins}
P.\ D.\ Higgins et al., \Journal{\PRD}{19}{1979}{731}.

\bibitem{barish}
S.\ J.\ Barish et al., \Journal{\PRD}{12}{1975}{1260}.

\bibitem{dao}
F.\ T.\ Dao et al., \Journal{\PRL}{30}{1973}{34}.

\bibitem{thomas}
A.\ W.\ Thomas and C. Boros, \Journal{\EPC}{9}{1999}{267}.


\bibitem{kopeliovich}
B.\ Kopeliovich, B.\ Povh and I.\ Potashnikova, \Journal{\ZPC}{73}{1996}{125}.

\bibitem{frank89}
L.\ L.\ Frankfurt, L.\ Mankiewicz and M.\ I.\ Strikman, \Journal{\ZPA}{334}{1989}{343}.

\bibitem{pumplin}
J.\ Pumplin, \Journal{\PRD}{7}{1973}{795}; \Journal{\PRD}{8}{1973}{2249}.

\bibitem{meissner}
T.\ Meissner, \Journal{\PRC}{52}{1995}{3386}.

\bibitem{holzwarth}
G.\ Holzwarth and R.\ Machleidt, \Journal{\PRC}{55}{1997}{1088}.

\bibitem{fries}
R. J.\ Fries and A.\ Sch\"afer, \Journal{\PRC}{57}{1998}{3470}.

\bibitem{nsss}
N.\ N.\ Nikolaev et al., 
\Journal{\PRD}{60}{1999}{014004}.

\bibitem{alesio}
U.\ D'Alesio and H.\ J.\ Pirner, \Journal{\EPA}{7}{2000}{109}.

\bibitem{nszak}
N.\ N.\ Nikolaev, J.\ Speth and B.\ G.\ Zakharov,
hep-ph/9708290 (1997).
















































\end{thebibliography}
\end{document}